\documentclass[traditabstract]{aa} 
\usepackage{txfonts}
\usepackage{graphicx}
\usepackage{natbib}
\bibpunct{(}{)}{;}{a}{}{,}
\begin{document}

\def\teff{$T_\mathrm{eff}$}
\def\logg{$\log g$}
\def\micro{$\xi$}
\def\kms{km s$^{-1}$}
\def\p{$\pm$}
\def\c{CoRoT}
\def\vsini{$v\sin i$}

   \title{A search for pulsations in the HgMn star HD 45975 with CoRoT photometry and ground-based spectroscopy\thanks{The CoRoT space mission was developed and is operated by the French space agency CNES, with participation of ESA's RSSD and Science Programmes, Austria, Belgium, Brazil, Germany, and Spain. This work is based on observations collected at La Silla and Paranal Observatories, ESO (Chile), with the HARPS and UVES spectrographs at the 3.6-m and very large telescopes, under programmes LP185.D-0056 and 287.D-5066. It is also based on observations made with the Mercator Telescope, operated on the island of La Palma by the Flemish Community, at the Spanish Observatorio del Roque de los Muchachos of the Instituto de Astrof\'{\i}sica de Canarias. Based on observations obtained with the HERMES spectrograph, which is supported by the Fund for Scientific Research of Flanders (FWO), Belgium, the Research Council of K.U. Leuven, Belgium, the Fonds National de la Recherche Scientifique (FNRS), Belgium, the Royal Observatory of Belgium, the Observatoire de Gen\`eve, Switzerland, and the Th\"uringer Landessternwarte Tautenburg, Germany.}}

   \titlerunning{A search for pulsations in the HgMn star HD 45975}
   \authorrunning{T. Morel et al.}

      \author{T. Morel
          \inst{1}
          \and
          M. Briquet
          \inst{1}
          \thanks{F.R.S.-FNRS Postdoctoral Researcher, Belgium}
          \and
          M. Auvergne
          \inst{2}
          \and
          G. Alecian
          \inst{3}
          \and
          S. Ghazaryan
          \inst{3,4}
          \and
          E. Niemczura
          \inst{5}
          \and
          L. Fossati
          \inst{6}
          \and
          H. Lehmann
          \inst{7}
          \and
          S. Hubrig
          \inst{8}
          \and
          C. Ulusoy
          \inst{9}
          \and
          Y. Damerdji
          \inst{10}
          \and
          M. Rainer
          \inst{11}
          \and
          E. Poretti
          \inst{11}
          \and
          F. Borsa
          \inst{11,12}
          \and
          M. Scardia
          \inst{11}
          \and
          V. S. Schmid
          \inst{13}
          \thanks{FWO Aspirant PhD Fellow, Belgium}
          \and
          H. Van Winckel
          \inst{13}
          \and
          K. De Smedt
          \inst{13}
           \and
          P. I. P\'apics
          \inst{13}
          \and
          J. F. Gameiro
          \inst{14}
          \and
          C. Waelkens
          \inst{13}
          \and
          M. Fagas
          \inst{15}
          \and
          K. Kami\'nski
          \inst{15}
          \and
          W. Dimitrov
          \inst{15}
          \and
          A. Baglin
          \inst{3}
          \and
          E. Michel
          \inst{3}
          \and
          L. Dumortier
          \inst{16}
          \and
          Y. Fr\'emat
          \inst{16}
          \and
          H. Hensberge
          \inst{16}
          \and
          A. Jorissen
          \inst{17}
          \and
          S. Van Eck
          \inst{17}
          } 

   \offprints{Thierry Morel, \email{morel@astro.ulg.ac.be}.}

   \institute{
   Institut d'Astrophysique et de G\'eophysique, Universit\'e de Li\`ege, All\'ee du 6 Ao\^ut, B\^at. B5c, 4000 Li\`ege, Belgium
   \and
   LESIA, Observatoire de Paris, CNRS, 5 place J. Janssen, 92190 Meudon, France 
   \and
   LUTH, Observatoire de Paris, CNRS, Universit\'e Paris Diderot, 5 place J. Janssen, 92190 Meudon, France 
   \and
   Byurakan Astrophysical Observatory, Aragatzotn Province, Byurakan, 0213, Armenia
   \and
   Astronomical Institute, Wroc\l aw University, Kopernika 11, 51-622 Wroc\l aw, Poland
   \and
   Argelander-Institut f\"ur Astronomie der Universit\"at Bonn, Auf dem H\"ugel 71, D-53121 Bonn, Germany
    \and
   Th\"uringer Landessternwarte Tautenburg (TLS), Sternwarte 5, 07778, Tautenburg, Germany 
   \and
   Leibniz-Institut f\"ur Astrophysik Potsdam (AIP), An der Sternwarte 16, 14482, Potsdam, Germany
   \and
   College of Graduate Studies, University of South Africa, PO Box 392, UNISA 0003, South Africa
   \and
   Centre de Recherche en Astronomie, Astrophysique et G\'eophysique (CRAAG), Route de l'Observatoire, BP 63 Bouzareah, Algiers, Algeria
   \and
   INAF -- Osservatorio Astronomico di Brera, via E. Bianchi 46, 23807 Merate (LC), Italy 
   \and
   Dipartimento di Scienza e Alta Tecnologia, Universit\`a dell'Insubria, Via Valleggio 11, 22100 Como (Italy)
   \and
   Instituut voor Sterrenkunde, K.U. Leuven, Celestijnenlaan 200D, 3001, Leuven, Belgium   
   \and
   Centro de Astrof\'{\i}sica e Faculdade de Ci\^encias da Universidade do Porto, Rua das Estrelas, 4150-762 Porto, Portugal 
   \and
   Astronomical Observatory Institute, Faculty of Physics, A. Mickiewicz University, S\l oneczna 36, 60-286 Pozna\'n, Poland
   \and
   Royal Observatory of Belgium, 3 avenue Circulaire, 1180 Brussels, Belgium  
   \and
   Institut d'Astronomie et d'Astrophysique, Universit\'e libre de Bruxelles, Boulevard du Triomphe CP 226, B-1050 Bruxelles, Belgium    
}

   \date{Received 16 July 2013; accepted 28 October 2013}
 
   \abstract{The existence of pulsations in HgMn stars is still being debated. To provide the first unambiguous observational detection of pulsations in this class of chemically peculiar objects, the bright star \object{HD 45975} was monitored for nearly two months by the CoRoT satellite. Independent analyses of the light curve provides evidence of monoperiodic variations with a frequency of 0.7572 d$^{-1}$ and a peak-to-peak amplitude of $\sim$2800 ppm. Multisite, ground-based spectroscopic observations overlapping the \c \ observations show the star to be a long-period, single-lined binary. Furthermore, with the notable exception of mercury, they reveal the same periodicity as in photometry in the line moments of chemical species exhibiting strong overabundances (e.g., Mn and Y). In contrast, lines of other elements do not show significant variations. As found in other HgMn stars, the pattern of variability consists in an absorption bump moving redwards across the line profiles. We argue that the photometric and spectroscopic changes are more consistent with an interpretation in terms of rotational modulation of spots at the stellar surface. In this framework, the existence of pulsations producing photometric variations above the $\sim$50 ppm level is unlikely in HD 45975. This provides strong constraints on the excitation/damping of pulsation modes in this HgMn star.}


   \maketitle
%

\section{Introduction}\label{sect_introduction}
A significant fraction of the stars of the upper main sequence conspicuously show abundance anomalies. One of the most spectacular examples are the so-called HgMn stars, which display extreme enhancements of manganese and mercury with overabundances that can reach up to 3 and 6 dex, respectively. Other elements (especially the heaviest ones) are also found to exhibit large deviations from the solar pattern and are either dramatically enhanced or depleted at the surface. In addition, isotopic ratios largely departing from the terrestrial values are commonplace \citep[e.g.,][]{hubrig99,dolk03}. The abundances \citep[e.g., Mn;][]{smith93} or isotopic ratios \citep[e.g., Hg;][]{smith97} of some elements appear to be correlated with the effective temperature. These late, main-sequence B stars have effective temperatures ranging from 10\,500 to 15\,000 K and masses roughly between 2.5 and 5 M$_{\odot}$ \citep[e.g.,][]{adelman03}. They are also often found in binary systems \citep[e.g.,][]{catanzaro04,scholler10,scholler12}. 

They are characterised by little evidence of turbulent motions in their atmospheres, weak stellar winds, and a generally slow rotation rate. Under such stable conditions and the lack of significant large-scale fluid motions in their interior, atomic diffusion is able to operate efficiently and to lead to the segregation of several chemical species as the result of a delicate balance between radiative forces and gravity (\citealt{michaud70} and, e.g., \citealt{alecian81} for HgMn stars). Diffusion provides a sound working hypothesis for modelling the bewildering array of elemental and isotopic abundances presented by HgMn stars, and these objects are considered as ideal laboratories for studying such hydrodynamical processes in detail. However, a number of key observational facts have still not been successfully modelled or may even challenge our very understanding of these objects. It is therefore important to provide observational evidence that other key physical processes are (or not) at work in HgMn stars. In particular, substantial effort has recently been directed towards detecting magnetic fields in these chemically peculiar objects \citep[e.g.,][]{wade06,makaganiuk11a,hubrig12,kochukhov13}.

The detection of pulsational instabilities in these stars would provide an opportunity to examine the excitation and damping of pulsation modes in stars with extreme manifestations of atomic diffusion processes. However, only a few theoretical and observational studies have been conducted so far. The models of \citet{turcotte_richard03} predict that HgMn stars should present a large iron opacity bump at about 2 $\times$ 10$^5$ K that would be very favourable to driving pulsations through the $\kappa$ mechanism. Calculations by \citet{alecian09} also suggest the excitation of non-radial $g$ modes as found in slowly pulsating B stars (SPBs). From the observational side, however, single HgMn stars do not show any evidence of large photometric changes \citep{adelman98}. This is rather unexpected considering the theoretical predictions, and it suggests that the models are still too crude and do not capture all the complexity of these objects, as indeed cautioned by \citet{turcotte_richard03,turcotte_richard05}, \citet{turcotte05}, and \citet{alecian09}.
 
Much tighter observational constraints have recently been provided by ultra-high precision photometric observations with the \c \ satellite \citep{baglin09}, which revealed for the first time periodic (with periods of 2.53 and 4.3 d) and low-amplitude (less than 1.6 mmag) changes compatible with pulsational instabilities in two faint ($V$ $>$ 12 mag) HgMn stars in the exoplanet fields \citep{alecian09}. As discussed below, the existence of spatially-extended abundance patches has been established from spectroscopy in a few HgMn stars \citep[e.g.,][]{adelman02}. However, arguments were presented that favour pulsations (as opposed to the rotational modulation of these structures) as being at the origin of the variability. Recent observations by the {\it Kepler} satellite \citep{borucki10} have shown similar changes in the HgMn star \object{KIC 612\,8830} (Sloan-like $g$ magnitude from KIC catalogue of 9.09 mag), namely nearly sinusoidal variations with a period of 4.84 d and a peak-to-peak amplitude of 3600 ppm \citep{balona11}. The variations observed were ascribed in this case to the rotational modulation of spots at the surface. A number of candidate variable HgMn stars have also been identified recently from observations with the {\it STEREO} spacecraft \citep{paunzen13}. Although the space-borne observations with \c \ and {\it Kepler} constitute by far the most sensitive search for pulsations in this class of objects, an unambiguous detection is unfortunately still lacking because the rotational modulation of abundance patches on the surface is also a viable interpretation for the changes observed both in terms of amplitudes and periodicities. 

These patches may not necessarily result from magnetic activity. \citet{kochukhovetal07} first detected mercury clouds with secular evolution on the surface of the bright HgMn star $\alpha$ And. Among the processes invoked by these authors to explain these observations, they considered the time-dependent atomic diffusion scenario proposed by \citet{alecian98}. Line-profile variations have been detected during the last decade in several HgMn stars and interpreted in terms of abundance spots in the photosphere \citep[e.g.,][]{adelman02,kochukhov05,hubrig06,briquet10}. These kinds of patches generally concern elements that have low abundances in normal stars (Sr, Y, Pt, Zr, and Hg) and which are suspected of being concentrated in high-altitude (above $\log \tau$ = --4) clouds. Such clouds in the upper atmosphere have also been diagnosed by propagating waves in rapidly oscillating Ap (roAp) stars \citep[e.g.,][]{kochukhov01,ryabchikova02}. High-altitude, uniform accumulation of Hg due to atomic diffusion was first proposed by \citet{michaud74} in their attempt to understand isotope anomalies of the mercury observed in some Ap stars. Since that pioneering work, high-altitude accumulations have often been considered and discussed in papers devoted to modelling atomic diffusion in atmospheres. Recently, \citet{alecian12} has proposed that abundance spots in some HgMn stars could be due to atomic diffusion in a weak (the diffusion velocity was computed for 100 Gauss in Alecian [\citeyear{alecian13}] suggesting a cloud location at $\log \tau = -4$)\footnote{In practice there is no theoretical lower threshold for the magnetic field intensity: the lower the magnetic field intensity, the higher the cloud.} and faintly organised magnetic field in their atmosphere, i.e., multipolar, varying on small spatial scales, and difficult to detect \citep[see also][]{alecian09}. Besides the possible effect of weak magnetic fields on the formation of high-altitude spots of some elements, the buildup of stratifications in the upper atmosphere is very complex since this process is non-linear, time-dependent, and possibly unstable \citep[see][]{alecian11}. This could provide an additional mechanism for disrupting a uniform stratification pattern and forming horizontal spot-like discontinuities of accumulations. 

In the course of our abundance study of the late B stars lying in the \c \ fields of view \citep{niemczura09},  we have identified a poorly studied B9 V star (HD 45975; HIP 31011) with strong overabundances of Hg and Mn that identify it as a member of the HgMn class. The abundances of several other elements were also found to strongly depart from the solar mixture. We present observations of this star acquired over two months through the seismology channel of \c. Rotational modulation and pulsational instabilities may lead to a different signature in the variations affecting the spectral features. Comparing the photometric and spectroscopic patterns of variability may therefore enable one to unambiguously establish the origin of the variations. Our observations of HD 45975 have been collected over a shorter timespan than was the case for the three other HgMn stars observed from space with \c \ or {\it Kepler}. However, the major advantage over previous investigations is that our target is easily amenable to monitoring from the ground of the spectral features thanks to its brightness ($V$ $\sim$ 7.5 mag). To lift the ambiguity between rotational modulation and pulsations, we have therefore conducted multisite spectroscopic observations from the ground in parallel with the \c \ observations. 

\section{Analysis of the \c \ data}\label{sect_corot_data}

HD 45975 was observed by the CoRoT satellite during the short run SRa04 lasting 53 days between 2011 October 6 and November 28 with an average time sampling of 32 seconds. 

\subsection{Processing of the data: ''imagettes''}\label{sect_imagettes}

\begin{figure}[t]
\centering
\includegraphics[width=7.5cm]{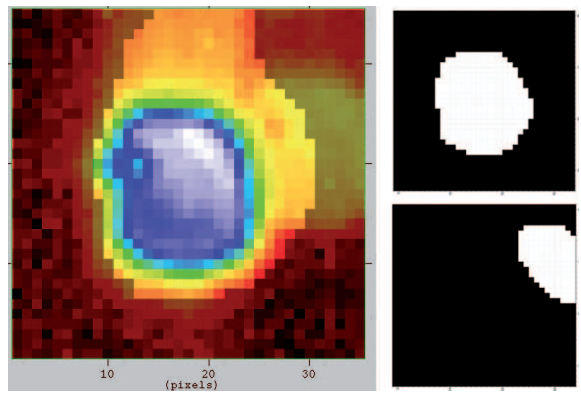}
\caption{\c \ ``imagettes'' of HD 45975. The left panel shows the first 2D image (35 $\times$ 35 pixels) of the series of 2193 consecutive ones analysed for this paper. The colour levels correspond to the number of collected photons (white to blue are for the brightest pixels, yellow, red to black for the faintest). The darker area around the target shows the  mask used for the light curve delivered as N2 data. North is to the right and east at the bottom. The image scale is 2.3$\arcsec$ pixel$^{-1}$. Masks 1 and 2 are shown in the upper and bottom right panels, respectively. }
\label{fig:imagette}
\end{figure}

Shortly after the beginning of the data acquisition, it appeared that the data may be affected by a faint contaminating star lying about 35$\arcsec$ to the north of HD 45975 (visible to the right of the main target in the left panel of Fig.~\ref{fig:imagette}). According to optical and near-infrared photometric data in the UCAC4 \citep{zacharias13}, NOMAD \citep{zacharias04}, and 2MASS \citep{cutri2003} catalogues, the visual companion is much redder and is fainter than HD 45975 by about four magnitudes in the $V$ band. Spectra of this source were obtained at Tautenburg (see Sect.~\ref{sect_observations}). This magnitude difference is consistent with the -- much less precise -- values derived from counts in spectra taken under similar conditions.\footnote{We estimated the stellar parameters of this source using the mean TLS spectrum, Kurucz atmosphere models, the 2010 version of the line analysis software MOOG \citep{sneden73}, and the line list of \citet{reddy03}. This star appears to be a moderately metal-poor, G subgiant: \teff \ $\sim$ 5700 K, \logg \ $\sim$ 3.5, and [Fe/H] $\sim$ --0.35. That it is less massive yet more evolved than HD 45975 suggests that they are not physically bound.} 

\begin{figure}[t]
\centering
\includegraphics[width=9.5cm]{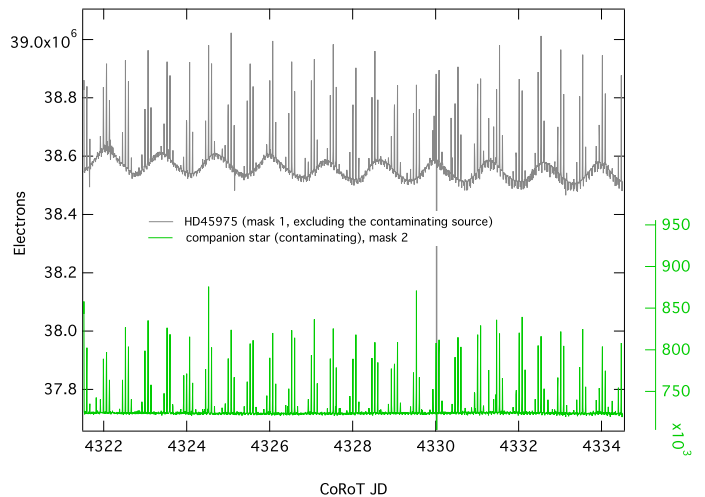}
\caption{Light curves (electrons vs. \c \ Julian days) of HD 45975 and the contaminant. Upper grey line: light curve using mask 1 (around HD 45975; $y$-axis to the left). Bottom green line: light curve using mask 2 (around the companion star; $y$-axis to the right, in the same units as the left axis, but on a different scale).}
\label{fig:LC_contaminant}
\end{figure}

The initial purpose of the small array centred on the target position (35 $\times$ 35 pixels acquired every 8 seconds) called ``imagettes'' in \c \ terminology is to control the PSF (point spread function) and the masking of each target. However, it is possible to use these ``imagettes'' to estimate the effect of a contamination because the electron counts are available for each pixel of these images. The asteroseismology channel has a defocussed PSF. As shown in Fig.~\ref{fig:imagette}, a small part of the contaminating source is not hidden by the numeric mask used to extract the light curve as an N2 data product. This mask is shown by the darker area around the target. To quantify this possible contamination, we considered a series of 2193 consecutive ``imagettes''  corresponding to 13 days of observations, with a 512 s binning. We measured the electron counts in these ``imagettes'', using two masks we defined for this specific case, one around the main target (mask 1) and the other one around the contaminating source (mask 2). These two masks are shown in Fig.~\ref{fig:imagette} (right panels) on a reduced scale. The light curves we obtained through these masks are shown in Fig.~\ref{fig:LC_contaminant}. These light curves were obtained from raw data (before the corrections were applied to produce the N2 data used in Sect.\ref{sect_light_curve}). The ratio of the counts for the pixels delimited by the two masks is only a rough estimate of the brightness difference between the two sources, but it agrees with the magnitude difference of $\sim$4 mag in the optical band determined above. The photometric variations that appear for HD 45975 are clearly absent from the light curve of the contaminating source. The light curve for HD 45975 is thus not affected by this companion. However, at this stage, and as for any other photometric observations limited by the instrument spatial resolution, one cannot exclude some other contamination from an unresolved source.

\subsection{Analysis of the light curve}\label{sect_light_curve}
\begin{figure}[t]
\centering
\includegraphics[width=9.0cm]{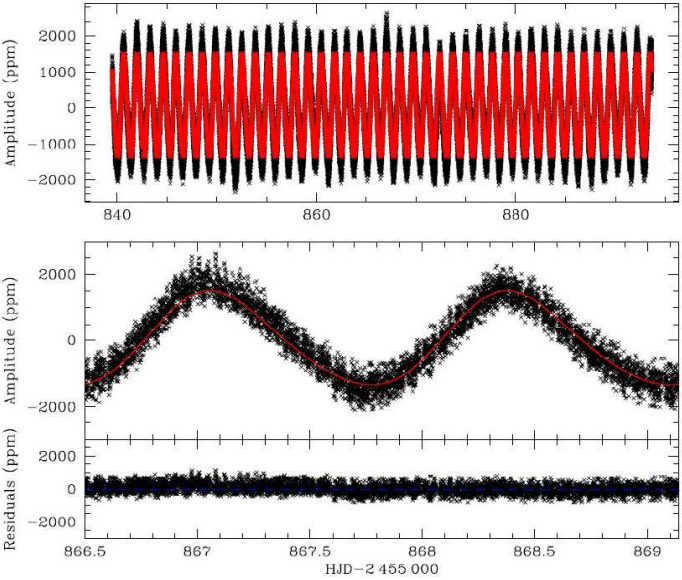}
\caption{\c \ light curve with the Fourier fit superposed ({\it red line}). The upper panel shows the full light curve, while the two bottom ones only show two cycles and the residuals (observations minus fit). The 1-$\sigma$ dispersion of the residuals is $\sim$300 ppm.}
\label{fig_light_curve}
\end{figure}

We first removed all flagged measurements, such as those suffering from hot pixels during the passage through the South Atlantic Anomaly. We then subtracted a downward trend considered to have an instrumental origin because similar behaviour is visible in all CoRoT light curves \citep[for details, see][]{auvergne09}. To remove the trend, we used different prescriptions, such as a linear approximation and a low-degree polynomial fit. It modified the low-frequency power excess that remained due to an imperfect detrending process (see below), but this did not affect the outcome of our analysis. The light curve consisting of 129\,615 data points, to which a second-order polynomial trend has been removed, is shown in the upper panel of Fig.~\ref{fig_light_curve}. One can clearly see variations with a period of about 1.3 day and a peak-to-peak amplitude of about 2800 ppm. Two cycles are displayed in the middle panel of this figure. The amplitude of the signal is comparable to those found for the three other HgMn stars observed with high-precision, space-based photometry \citep{alecian09,balona11}. 

\begin{figure}[t]
\centering
\includegraphics[width=9.0cm]{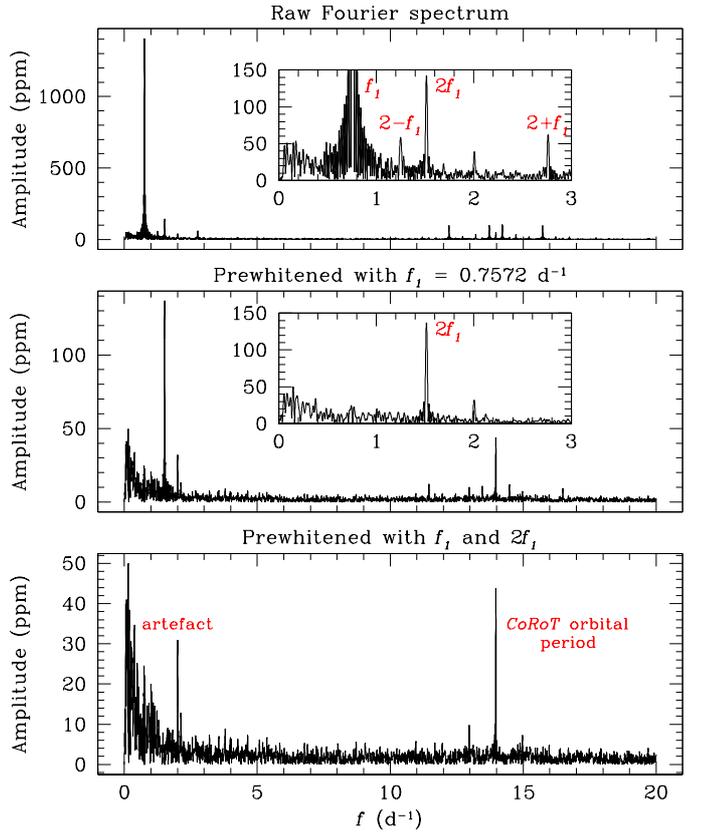}
\caption{Raw Fourier spectrum ({\it upper panel}) and after prewhitening with $f_1$ ({\it middle panel}) and with $f_1$ and 2$f_1$ ({\it bottom panel}).}
\label{fig_fourier_corot}
\end{figure}

The Fourier spectrum of the CoRoT light curve, computed with the Period04 software \citep{lenz05}, is shown in the upper panel of Fig.~\ref{fig_fourier_corot}. A dominant frequency at $f_1$ = 0.7572~d$^{-1}$ (period of about 1.321 d), together with its harmonics 2$f_1$, as well as the aliases 2--$f_1$ and 2+$f_1$, is visible. The aliases are due to the presence of a peak at 2~d$^{-1}$, which is seen in all CoRoT light curves and is thus not intrinsic to the star. \citet{auvergne09} list some of the most important spurious sources of signal in CoRoT observations. One of them is Earth infrared emissivity, which has a nearly 24-hour cycle because of the night and day cycle of the Earth. This could possibly be the cause of this 2~d$^{-1}$ feature. The Fourier spectra after prewhitening with $f_1$ and then with $f_1$ and 2$f_1$ are displayed in the middle and bottom panels of Fig.~\ref{fig_fourier_corot}, respectively. Other peaks stand out, but they are not attributed to the star. Besides the peak at 2~d$^{-1}$ discussed above, there are peaks linked to the orbital frequency of the CoRoT satellite, $f_{\rm orb}$ = 13.97~d$^{-1}$. In the high-frequency domain above 20~d$^{-1}$, harmonics of the satellite orbital frequency up to 8$f_{\rm orb}$ are seen. After prewhitening with these artefacts, there is no evidence of additional frequencies that can be ascribed to the star down to an amplitude lower than $\sim$50 ppm. A harmonic fit with only $f_1$ and 2$f_1$ is superimposed on the data in Fig.~\ref{fig_light_curve}.

An independent analysis using a different method for the treatment of the light curve (procedure {\sc ColiGcor}; see \citealt{ghazaryan13}) led to the same conclusions regarding the monoperiodic nature of the variations and the value of the fundamental frequency.

\section{Analysis of the ground-based observations}\label{sect_ground_based}

\subsection{Observations and data reduction}\label{sect_observations}
The atmosphere of HgMn stars is highly dynamical and spots at the surface may evolve relatively rapidly \citep{kochukhovetal07,briquet10,korhonen13}. It is therefore important to secure the spectroscopic data as closely as possible to (and, as far as possible, simultaneously with) the \c \ observations. A coordinated, multisite campaign involving seven telescopes worldwide was therefore organised to support the \c \ observations. Table \ref{tab_observations} summarises the telescopes and instruments used. This is, to our knowledge, the most extensive spectroscopic monitoring of a HgMn star ever attempted. 

\begin{table*}
\caption{Details of the ground-based, spectroscopic observations.}
\label{tab_observations}
\centering
\begin{tabular}{llrcccc}
\hline\hline
Instrument               & Telescope                                             & $N$ & $R$                       & $\Delta \lambda$ (\AA) & S/N     & Date of observation\\
                         &                                                       &     &                           &                        &         & (HJD -- 2\,450\,000)\\
\hline
FEROS                    & ESO, La Silla (2.2m)                                  &  1  &  48\,000                  & 3750--8530             & 220      & 2655.57\\
HERMES                   & Mercator (1.2 m)                                      & 39  &  85\,000                  & 3770--9000             & 115--220 & 5218.45--5889.63\\
Coud\'e                  & Th\"uringer Landessternwarte (TLS) Tautenburg (2.0 m) & 40  &  63\,000                  & 4720--7360             & 65--195 & 5840.64--5939.38\\
Echelle                  & Pozna\'n Spectroscopic Telescope (PST)                & 21  &  35\,000                  & 4300--7510             & 10--45   & 5847.60--5895.60\\
UVES                     & ESO, Paranal (VLT)                                    & 16  &  80\,000\tablefootmark{a} & 3290--4520             & 215--360 & 5855.85--5950.55\\
                         &                                                       & 16  & 110\,000\tablefootmark{a} & 5690--7520             & ...      & 5855.85--5950.55\\  
                         &                                                       & 16  &  95\,000\tablefootmark{a} & 7660--9460             & ...      & 5855.85--5950.55\\  
HARPS                    & La Silla (3.6 m)                                      & 77  &  80\,000                  & 3780--6910             & 175--355 & 5913.61--5937.84\\
GIRAFFE\tablefootmark{b} & South African Astronomical Observatory, SAAO (1.9 m)  &  8  &  39\,000                  & 3840--5850             & 30--120  & 5925.38--5930.51\\
SARG                     & TNG (3.6 m)                                           & 65  &  46\,000                  & 3640--4310             & ...      & 5967.34--5980.57\\
                         &                                                       & 65  &  46\,000                  & 4360--5140             & 155--270 & 5967.34--5980.57\\
\hline 
\end{tabular}
\tablefoot{$N$ is the total number of spectra collected, $R$ is the mean resolving power, and $\Delta \lambda$ is the wavelength coverage. As discussed in Sect.~\ref{sect_lpv}, the S/N has been determined in a homogeneous way and is based on the 1-$\sigma$ dispersion measured in continuum regions at $\sim$4500 \AA \ ($\sim$4965 \AA \ for the TLS spectra). The HERMES and HARPS observations were carried out in the HRF and EGGS modes, respectively. \tablefootmark{a}{Estimated based on figures quoted in UVES User's manual between the resolving power and the slit width (here 0.4 and 0.3\arcsec for the blue and red arms, respectively)}; \tablefootmark{b}{The Grating Instrument for Radiation Analysis with a Fibre Fed \' Echelle (GIRAFFE) instrument is a replica of the MUSICOS spectrograph developed at Observatoire de Paris, Meudon \citep{baudrand92}.}
}
\end{table*}

The HERMES, TLS, PST, and UVES data overlapped the \c \ observations, while the HARPS, GIRAFFE, and SARG ones were taken slightly after the end of the space observations. The HARPS spectra were obtained in the framework of the ESO large programme LP185.D-0056 \citep{poretti13}.  We also considered one archive FEROS spectrum obtained much earlier and retrieved from the GAUDI database \citep{solano05}. 
 
The HERMES data were reduced using the instrument pipeline \citep{raskin11}. For the TLS spectra, standard ESO-MIDAS packages, and our own routine for determining the instrumental zero point in radial velocity that uses the telluric O$_2$ lines were employed. The PST spectra were reduced using a dedicated reduction pipeline \citep[see][]{baranowski09}. The reduction of the UVES spectra was performed under the ESO-REFLEX\footnote{See {\tt http://www.eso.org/sci/software/reflex/.}} environment in interactive mode to allow for an optimal choice of the parameters. For the HARPS spectra, dedicated tools developed at Brera observatory were used \citep{poretti13}. The GIRAFFE spectra were reduced using the {\tt SPEC2} package described by \citet{balona96}. For the SARG spectra, the usual reduction steps for \'echelle spectra (i.e., bias subtraction, flat-field correction, removal of scattered light, order extraction, wavelength calibration, and merging of the orders) were carried out using the standard tasks implemented in the IRAF\footnote{{\tt IRAF} is distributed by the National Optical Astronomy Observatories, operated by the Association of Universities for Research in Astronomy, Inc., under cooperative agreement with the National Science Foundation.} software. 

The PST and GIRAFFE spectra are of much lower quality (see Table \ref{tab_observations}) and will not be discussed further. The TLS spectra only cover a few of the diagnostic lines used in the following and are also of lower quality than those obtained with HERMES, UVES, HARPS, and SARG. They therefore do not provide additional information. However, these data allowed us to characterise the nature of a contaminating source affecting the \c \ observations (see footnote 1 in Sect.~\ref{sect_imagettes}).

\subsection{Fundamental parameters and chemical composition}\label{sect_abundance_analysis}
Two abundance analyses (referred to in the following as ``first'' and ``second'' analyses) assuming local thermodynamic equilibrium (LTE) have been independently conducted employing different diagnostics, tools (e.g., codes, model atmospheres) and atomic data.

We first used line-blanketed, solar-metallicity atmospheric models computed with ATLAS9 and synthetic spectra calculated with SYNTHE \citep{kurucz1993}. The line identification was based on a spectral atlas of the HgMn star \object{HD 175640} \citep{castelli04}.\footnote{See also: {\tt http://wwwuser.oat.ts.astro.it/castelli/ \linebreak hd175640/tab3040-10000.html}} This was also the source of the atomic data. The wavelength intervals investigated were 3900--8500 \AA \ for HERMES and 3315--3650 \AA \ for UVES. We found \teff \ = 12\,500\p500 K and \logg \ = 4.2\p0.2 by finding the best match between the observed Balmer line profiles and a library of synthetic spectra computed for solar metallicity. We also determined \teff \ and the microturbulence, \micro, by requiring that the abundances derived from line-profile fitting of a set of 65 \ion{Fe}{ii} lines are independent of the lower excitation potential and strength. The surface gravity was frozen to the value derived above. We found \teff \ = 12\,700 K and \micro \ = 0.5\p0.5 \kms. These two values are supported by the analysis of other metal ions (\ion{Ti}{ii}, \ion{Cr}{ii}, and \ion{Mn}{ii}). We closely followed the methodology presented by \citet{niemczura09} to determine the chemical abundances.

In the second approach, we simultaneously used the hydrogen lines measured in the HERMES spectra and the spectral energy distribution (SED) to estimate \teff \ and \logg. We employed {\sc LLmodels} model atmospheres \citep{shulyak2004} computed for an appropriate choice of abundances. As a source of atomic line parameters for opacity calculations and abundance determination, we used the {\sc VALD} database \citep{piskunov95,kupka99,ryabchikova99}. To analyse the hydrogen lines, we fitted synthetic spectra calculated with {\sc SYNTH3} \citep{kochukhov07} iteratively taking into account the individual abundances in the model calculation. Also considering the \teff \ and \logg \ values estimated from the SED analysis (see below), we finally adopted \teff \ = 12\,250\p500 K and \logg \ = 4.2\p0.1. The \ion{Ti}{ii} and \ion{Fe}{ii} excitation equilibrium further confirmed the adopted \teff \ value.

The equivalent widths (EWs) of several weakly blended lines were analysed with a modified version \citep{tsymbal96} of the {\sc WIDTH9} code \citep{kurucz1993}. By requiring no dependence for \ion{Ti}{ii} and \ion{Fe}{ii} between the abundances and the EWs, we obtained \micro \ = 1.0\p1.0 \kms. The abundances were derived using synthetic spectra and the tools described in \citet{fossati2007}. We made use of the average HERMES and UVES spectra for the visible (between the Balmer jump and the hydrogen Paschen series) and blue (beyond the Balmer jump) regions, respectively. Several Mn lines show flat cores indicative of the presence of hyperfine structure (HFS). Although we could not systematically take HFS into account owing to the unavailability of atomic data, this was done for 4 lines using the data of \citet{holt1999}. For \ion{Ga}{ii} $\lambda$6334, we adopted the atomic parameters of \citet{joensson2006}. We checked for overabundances of other elements typically present in HgMn stars, such as Ne, P, Ar, Xe, and Cl \citep[e.g.,][]{dworetsky08,fossati2011a}. However, no lines of these species are confidently detected and increasing the abundances starting from the solar value leads to a worse fit to the observations.

\begin{figure*}[t]
\centering
\sidecaption
\includegraphics[width=11.0cm,clip]{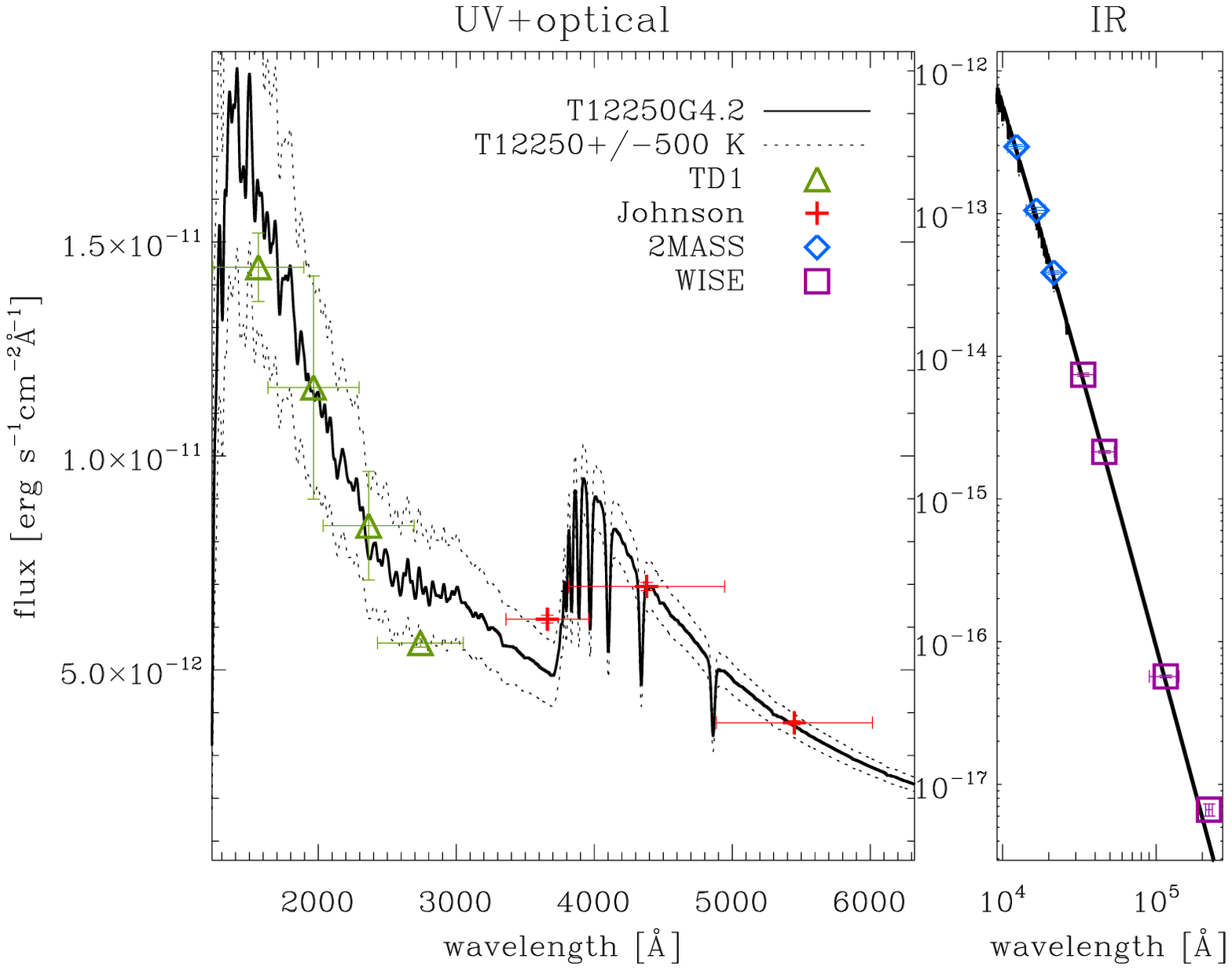}
\caption{Comparison between {\sc LLmodels} theoretical fluxes ({\it full line}) and TD1 ({\it triangles}; \citealt{thompson1995}), Johnson ({\it crosses}; \citealt{mermilliod1994}), 2MASS ({\it diamonds}; \citealt{cutri2003}), and WISE ({\it squares}; \citealt{cutri2012}) photometry converted to physical units. We adopted the calibrations of \citet{thompson1995}, \citet{bessell1998}, \citet{bliek1996}, and \citet{wright2010}, respectively. The large deviation of the TD1 photometric point at $\sim$2740 \AA \ is a well-known problem most likely of instrumental origin \citep[e.g.,][]{malagnini83,kjaergaard84}, and this measurement was not taken into account during the fitting. All the horizontal error bars correspond to the full width at half-maximum of the filter transmission curves. A convolution was applied to the theoretical fluxes for visualisation purposes. Dotted lines show the effect of a 500 K temperature variation in the theoretical fluxes. Note the $x$-axis logarithmic scale of the right panel.}
\label{fig_sed}
\end{figure*}

Figure \ref{fig_sed} shows the fit of the synthetic fluxes, calculated with the fundamental parameters and abundances derived above, to the observed photometry. Our SED analysis was performed fitting simultaneously \teff, \logg, the stellar radius, and the interstellar extinction. Adopting a reddening $E$($B$--$V$) variable within the 0.00--0.05 range \citep{amores2005}, we obtained as best-fitting parameters: \teff \ = 12\,000\p250 K, \logg \ = 4.2\p0.3, $R/R_{\odot}$ = 1.8\p0.5, and $E$($B$--$V$) = 0.035\p0.010 mag. The radius was obtained by fitting the synthetic fluxes to the flux-calibrated photometry using the Hipparcos distance of 203\p21 pc \citep{leeuwen2007} whose uncertainty constitutes the major source of error. Our value is not very dependent on the reddening because we make use of data covering a wide wavelength range, which includes the near- and mid-IR domains. We compared the observed Johnson and 2MASS photometric colours to synthetic ones calculated with the synthetic fluxes and extinction used in Fig.~\ref{fig_sed} \citep[see][]{fossati2011b}. The agreement is below the 1$\sigma$ level, confirming the validity of our atmospheric parameters. 

Table \ref{tab_abundances} summarises the results of the two abundance analyses. The parameters agree closely, and we consider \teff \ = 12\,375\p500 K and \logg \ = 4.2\p0.1 as final values. We note that such an effective temperature is not compatible with the B9 classification assigned in the SIMBAD database and is more typical of a B7 or B8 dwarf. This misclassification is likely due to the abnormal weakness of the \ion{He}{i} lines \citep{dworetsky1976}. Line-profile fitting or computation of the Fourier transform \citep[e.g.,][]{reiners2001} for a set of unblended metal lines of different ions leads to \vsini \ = 61\p2 \kms.

\begin{table}[t]
\caption{LTE atmospheric abundances of HD 45975 derived using the two independent analyses.}
\centering
\label{tab_abundances}
\begin{tabular}{l|cr|cr|c}
\hline\hline
          & \multicolumn{4}{|c|}{HD 45975} & Sun \\      
          & \multicolumn{2}{|c|}{First analysis} & \multicolumn{2}{|c|}{Second analysis} & \\      
\hline 
Element   & $\log (\cal{N}/\cal{N}_{\rm tot})$ & $n$ & $\log (\cal{N}/\cal{N}_{\rm tot})$ & $n$ & $\log (\cal{N}/\cal{N}_{\rm tot})$ \\
\hline
He        &                &    &  $-$1.40\p0.23  &  2 &  $-$1.11 \\
C         &  $-$3.77\p0.32 &  1 &  $-$3.78\p0.32  &  1 &  $-$3.61 \\
O         &  $-$3.35\p0.22 &  4 &  $-$3.47\p0.09  &  2 &  $-$3.35 \\
Mg        &  $-$4.93\p0.20 & 10 &  $-$4.74\p0.17  &  4 &  $-$4.44 \\
Al        &  $-$6.55\p0.05 &  2 &  $-$6.40\p0.21  &  1 &  $-$5.59 \\
Si        &  $-$4.75\p0.21 & 16 &  $-$4.87\p0.17  &  4 &  $-$4.53 \\
S         &  $-$5.13\p0.34 &  7 &  $-$4.75\p0.20  &  4 &  $-$4.92 \\
Ca        &  $-$5.76\p0.29 &  1 &  $-$5.61\p0.29  &  1 &  $-$5.70 \\
Sc        &  $-$9.07\p0.29 &  1 &                 &    &  $-$8.89 \\
Ti        &  $-$5.87\p0.26 & 26 &  $-$5.88\p0.24  & 35 &  $-$7.09 \\
Cr        &  $-$5.65\p0.26 & 29 &  $-$5.44\p0.18  & 36 &  $-$6.40 \\
Mn        &  $-$4.60\p0.28 & 73 &  $-$4.37\p0.22  & 63 &  $-$6.61 \\
Mn (HFS)  &                &    &  $-$4.55\p0.12  &  4 &  $-$6.61 \\
Fe        &  $-$4.86\p0.18 & 65 &  $-$4.80\p0.17  & 46 &  $-$4.54 \\
Ni        &  $-$6.82\p0.18 &  4 &                 &    &  $-$5.82 \\  
Ga\tablefootmark{a}        &  $-$5.94\p0.27 &  1 &                 &    &  $-$9.00 \\
Ga (HFS)\tablefootmark{a}  &                &    &  $-$5.30\p0.27  &  1 &  $-$9.00 \\
Sr        &  $-$8.83\p0.35 &  1 &  $-$9.76\p0.35  &  1 &  $-$9.17 \\
Y         &  $-$6.33\p0.27 & 17 &  $-$6.51\p0.23  & 19 &  $-$9.83 \\
Hg        &  $-$5.36\p0.38 &  4 &  $-$4.87\p0.17  &  4 & $-$10.87 \\
\hline
\end{tabular}
\tablefoot{
${\cal N}$ is the number density of the species and $n$ is the number of measured lines. The Mn and Ga abundances are given taking into account or ignoring HFS effects. For comparison purposes, the last column lists the (rescaled) solar abundances derived from 3D hydrodynamical model atmospheres
\citep{asplund2009}. \tablefootmark{a}{The gallium abundance is particularly uncertain because it depends on the treatment of the blended \ion{Ne}{i} $\lambda$6334 line.}}
\end{table}

\begin{figure}[h]
\centering
\includegraphics[width=9.0cm]{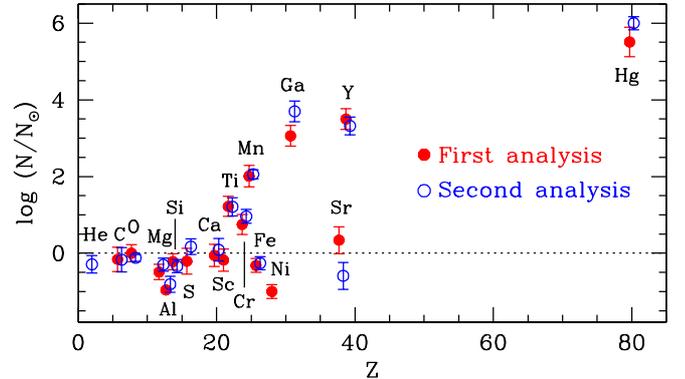}
\caption{Abundance pattern of HD 45975 relative to that of the Sun ({\it filled and open symbols:} first and second analyses, respectively). For the second analysis, HFS is taken into account for the Mn abundance.}
\label{fig_abundances}
\end{figure}

The errors in the abundances arising from uncertainties in the adopted stellar parameters were estimated by repeating the calculations after varying in turn one of the three parameters (\teff, \logg, and \micro) by its 1-$\sigma$ uncertainty (500 K, 0.1 dex, and 0.5 \kms, respectively) while keeping the other two unchanged. These three sources of errors and the line-to-line scatter (a conservative estimate of 0.2 dex was assumed in case the abundance was based on a single line) were quadratically summed to obtain the final abundance uncertainties quoted in Table \ref{tab_abundances}. That the abundances derived by the two independent analyses agree within the errors for the vast majority of elements (see Fig.~\ref{fig_abundances}) suggests that these uncertainties are representative of the accuracy of our results.

The unusually high abundances previously reported by \citet{niemczura09} are confirmed. With respect to solar, HD 45975 exhibits strong excesses of various elements, as commonly observed in HgMn stars \citep[e.g.,][]{adelman01}: Ti and Cr ($\sim$1 dex), Mn ($\sim$2 dex), Ga and Y ($\sim$3.5 dex), and Hg ($\sim$6 dex). 

One caveat of our analysis is the neglect of departures from LTE. Non-LTE corrections for late B dwarfs have been computed by \citet{hempel03} for a number of elements. In the relevant temperature range (i.e., ignoring the dependence with other parameters), the following typical corrections (non-LTE minus LTE abundances) were obtained: C ($\sim$+0.15 dex), O ($\sim$--0.6 dex), Mg ($\sim$--0.05 dex), Si ($\sim$--0.2 dex), Ca ($\sim$+0.2 dex), Fe ($\sim$--0.05 dex), and Sr ($\sim$+0.3 dex). It should be kept in mind that these calculations are based on modelling a set of spectral lines that are not necessarily those used in our abundance analysis (e.g., the \ion{O}{i} triplet at $\sim$7770 \AA). Furthermore, they may not apply to the peculiar conditions prevailing in the photospheres of HgMn stars. With these limitations in mind, it is nonetheless very likely that the non-LTE corrections are substantial for some elements. A detailed investigation is, however, beyond the scope of this paper. 

We also ignored the possible existence of a  vertical stratification of the various chemical elements in the photosphere. Although there is no compelling observational evidence for this phenomenon in HgMn stars \citep{savanov03,thiam10,makaganiuk12}, an accumulation at different layers is supported by theoretical calculations (Sect.~\ref{sect_discussion}). However, even if present and taken into account in HD 45975, it is likely that our atmospheric parameters would remain within the uncertainties \citep[see the analysis of an early Ap star by][]{pandey11}.

\begin{figure}[h]
\centering
\includegraphics[width=8.5cm]{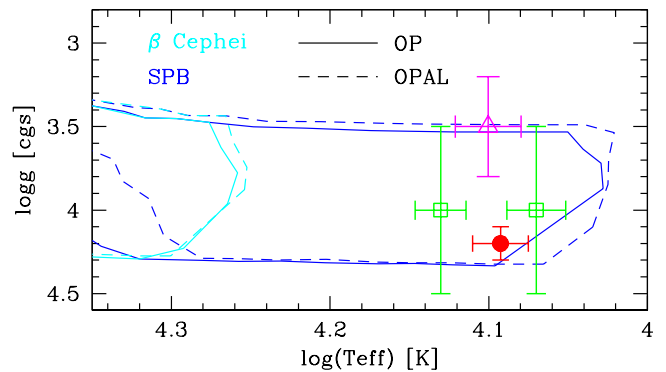}
\caption{Position of HD 45975 ({\it filled circle}) compared to instability strips at solar metallicity for $\beta$ Cephei and SPB stars \citep{miglio07}. The domains of instability are shown for computations based on OP or OPAL opacities. The position of the other HgMn stars observed with \c \ \citep[{\it open squares};][]{alecian09} and {\it Kepler} \citep[{\it open triangle};][]{balona11} is also shown. The determination of the parameters for \object{KIC 612\,8830} is discussed by \citet{catanzaro10}.}
\label{fig_instability_strips}
\end{figure}

\begin{figure}[h]
\centering
\includegraphics[width=7.5cm]{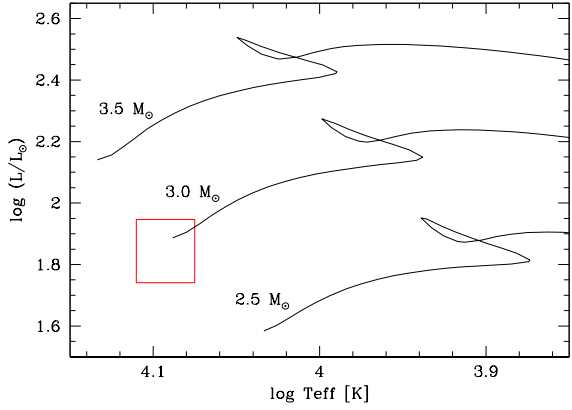}
\caption{Position of HD 45975 in the Hertzsprung-Russell (HR) diagram assuming the final parameters (delimited by a 1-$\sigma$ error box). Padova evolutionary tracks at solar metallicity \citep{bertelli09} for masses of 2.5, 3.0, and 3.5 M$_\odot$ are overplotted. The radius of the 3.0 M$_\odot$ model on the ZAMS is 1.97 R$_\odot$, while the \logg \ is 4.33.}
\label{fig_HR}
\end{figure}

With our parameters, the star appears to be situated in the instability strip for SPB-like oscillations (Fig.~\ref{fig_instability_strips}). It should, however, be noted that neither rotation nor diffusion is implemented in these models. Using the visual magnitude, the bolometric corrections from \citet{flower96}, and the reddening and distance derived above, we inferred $\log \,$($L/L_{\sun}$) = 1.84\p0.11. As shown in Fig.~\ref{fig_HR}, a comparison with evolutionary tracks \citep{bertelli09} shows that our \teff, \logg, radius, and luminosity estimates are all compatible with a $\sim$3 M$_{\odot}$ star that is very close to the zero-age main sequence (ZAMS). HD 45975 lies in the \object{NGC 2232} field of view, which is at a distance of about 352 pc \citep{leeuwen2009}. It is unlikely that the star is a cluster member: its position with respect to evolutionary tracks would indicate a $\sim$3.5 M$_{\odot}$ star that has substantially evolved off the ZAMS, and with a \logg \ significantly lower than our estimate. This conclusion agrees with the peculiar proper motion of HD 45975 compared to the cluster stars \citep{leeuwen2009}.

\subsection{Line-profile variations and binarity}\label{sect_lpv}

A search for short-term line-profile variations was performed using the most extensive and highest quality time series (namely UVES, HARPS, HERMES, and SARG). The data discussed here were obtained over a relatively short time interval, and the possible superposition of secular changes arising from the dynamical evolution of the stellar atmosphere in case of spots at the surface \citep[e.g.,][]{kochukhovetal07,hubrig10,korhonen13} should not be a concern when considering a single dataset. However, it cannot be ruled out that a lack of coherency is encountered when comparing one time series to another (for instance, the HERMES and SARG datasets are separated by $\sim$4 months). Neither four HERMES spectra taken much earlier than the \c \ observations to explore the nature of the HgMn stars identified in the fields of view of the satellite \citep{niemczura09} nor the two HARPS ones taken much later were considered because of large differences in radial velocity arising from binary motion (see below). 

First, a set of unblended lines of some key elements was selected from an atlas of the HgMn star \object{HD 175640} \citep{castelli04}. The lines investigated for line-profile variability were \ion{He}{i} $\lambda$4713, \ion{Hg}{ii} $\lambda$3984, \ion{Si}{ii} $\lambda$4131, \ion{Mn}{ii} $\lambda$4137, \ion{Mg}{ii} $\lambda$4481, \ion{Cr}{ii} $\lambda$4559, \ion{Ti}{ii} $\lambda$3913, 4564, 4572, \ion{Y}{ii} $\lambda$4900, 5663, and \ion{Fe}{ii} $\lambda$5169. No gallium lines are unfortunately suitable because of blending with other atomic and/or telluric features. No strontium lines can be used either because of their extreme weakness (the Sr abundance is near or below solar in HD 45975; Table \ref{tab_abundances}). In general, as was the case when estimating the abundances, the high rotation rate prevents any analysis of elements with intrinsically weak spectral features. To ensure consistency, all profiles for a given dataset were normalised using the same continuum regions and fitting functions.

We first investigated changes in the global line-profile properties by calculating the observed line moments \citep[see, e.g.,][]{aerts92} with FAMIAS \citep{zima08}. We examined changes in the zeroth (EW expressed in velocity units), first (centroid), second (variance), and third (skewness) moments of the line profile (referred to as M$_0$, M$_1$, M$_2$, and M$_3$ in the following). The three last moments are normalised by M$_0$. The moments were computed within fixed velocity intervals encompassing the line profile. To estimate the S/N, we also used FAMIAS and calculated the standard deviation in continuum regions in close vicinity to the line of interest for each exposure through an iterative sigma-clipping algorithm. A discrete Fourier transform was computed with a frequency step of (20$\Delta T$)$^{-1}$ (where $\Delta T$ is the total time span of the data) to detect periodic signals in these four quantities. 

\begin{figure}[t]
\centering
\includegraphics[width=7.5cm]{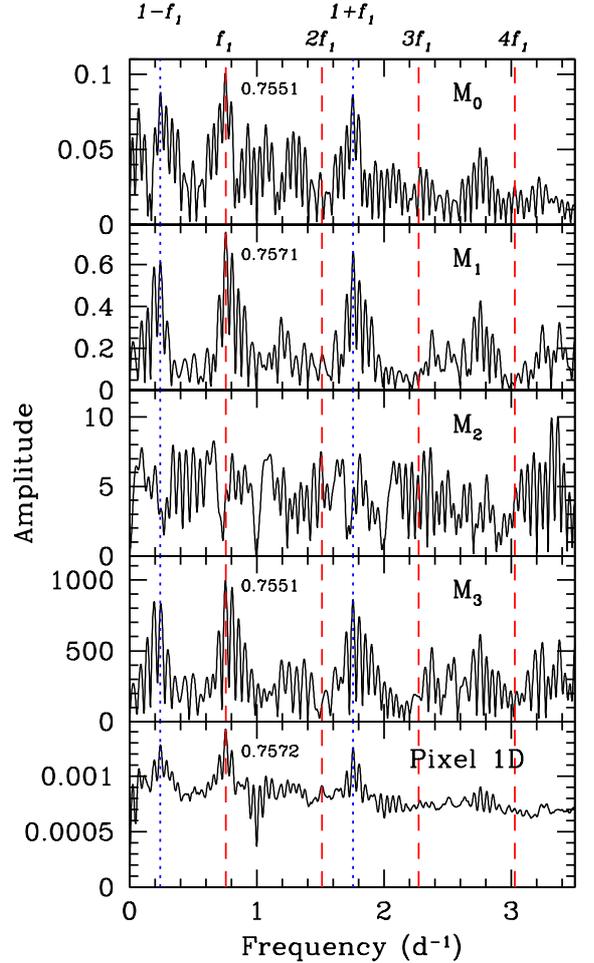}
\caption{Fourier spectra for \ion{Y}{ii} $\lambda$4900 of the four line moments ({\it four upper panels}) and mean Fourier spectrum for the pixel-to-pixel variations across the line profile ({\it bottom panel}). The frequency detected in the \c \ data (and its harmonics) are indicated as dashed lines, while the dotted lines correspond to 1--$f_1$ and 1+$f_1$. The fundamental frequency is indicated when a periodic signal is detected.}
\label{fig_fourier_Y_II_4900}
\end{figure}

This led to negative results for the HERMES, SARG, and UVES data. For the HARPS data, however, a prominent peak in the power spectrum of M$_0$ and/or M$_1$ was repeatedly found at a frequency close to the one detected in the \c \ data. This was also occasionally the case for M$_3$ (but not for M$_2$). Figure \ref{fig_fourier_Y_II_4900} shows representative examples in the case of \ion{Y}{ii} $\lambda$4900. Unlike the \c \ light curve, there is no significant power in the harmonics. The lack of detection for the other datasets is likely due to the much lower quality of the HERMES spectra, to the poor phase sampling of the SARG observations, and to the much lower number of spectra collected with UVES (16 against 75). We assume in the following that all these frequencies can be, within the errors, identified with $f_1$ (Sect.~\ref{sect_light_curve}). 

\begin{figure*}
\begin{minipage}[t]{0.48\textwidth}
\includegraphics[width=\textwidth]{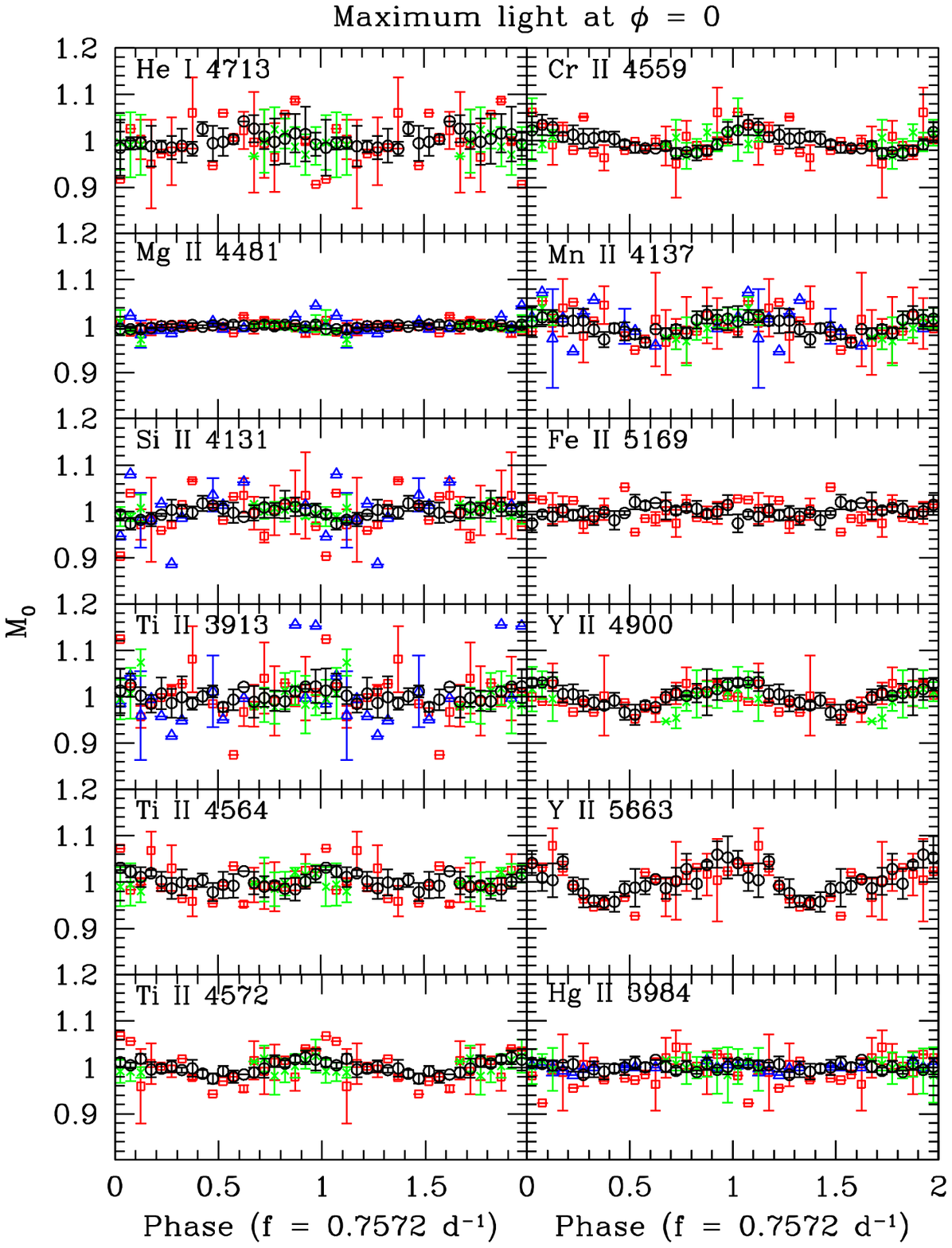}
\end{minipage}
\begin{minipage}[t]{0.48\textwidth}
\includegraphics[width=\textwidth]{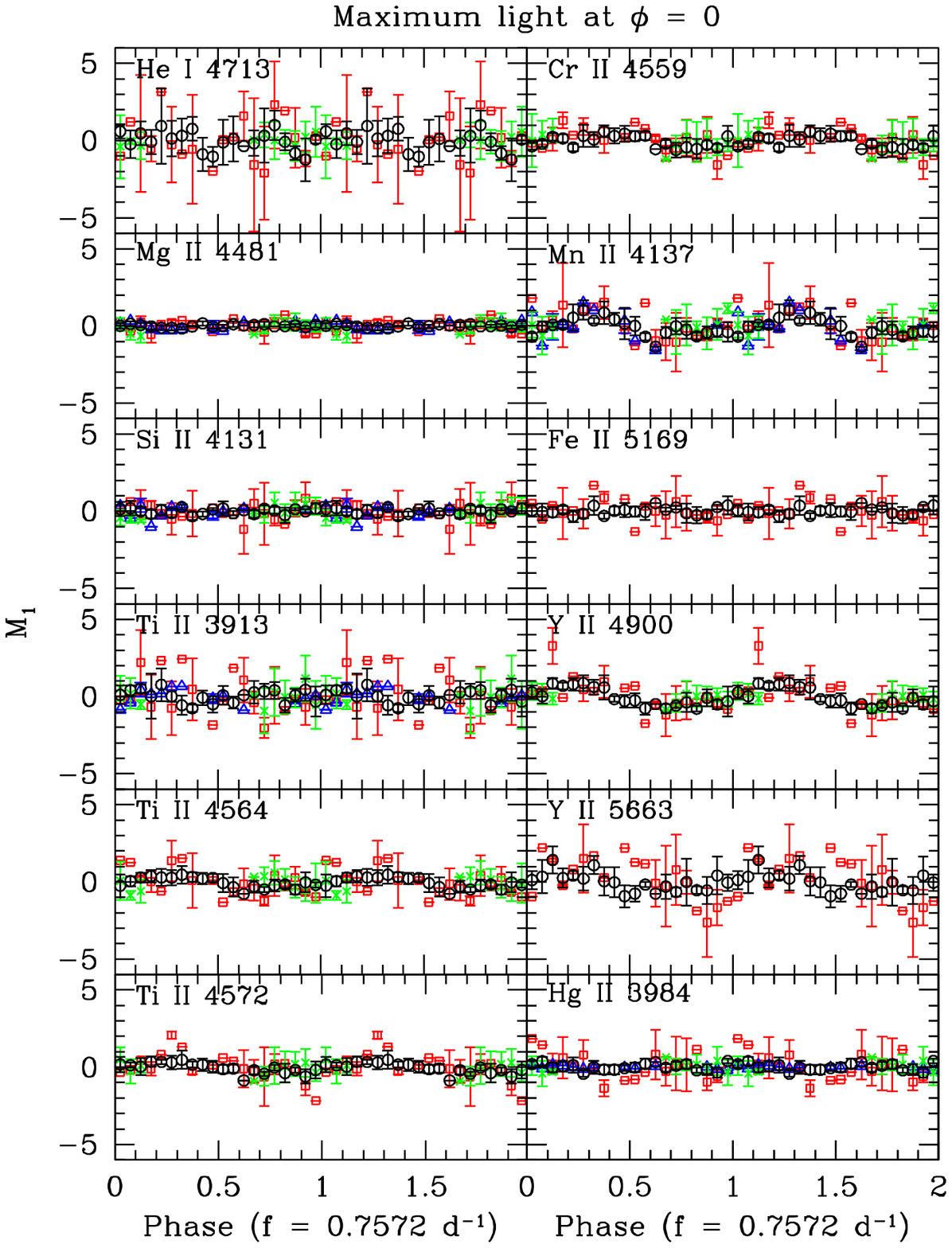}
\end{minipage}
\begin{minipage}[b]{0.48\textwidth}
\includegraphics[width=\textwidth]{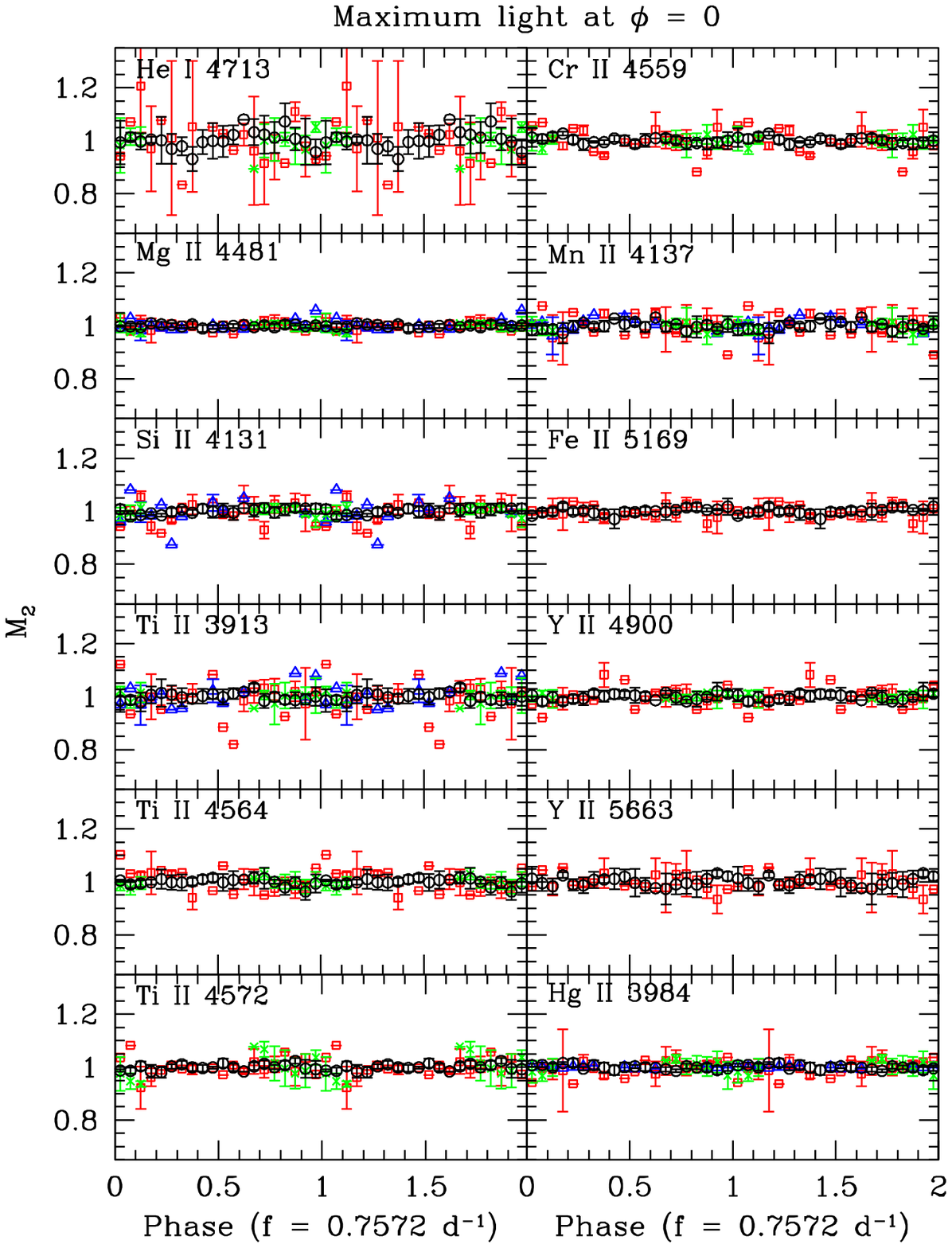}
\end{minipage}
\begin{minipage}[b]{0.48\textwidth}
\includegraphics[width=\textwidth]{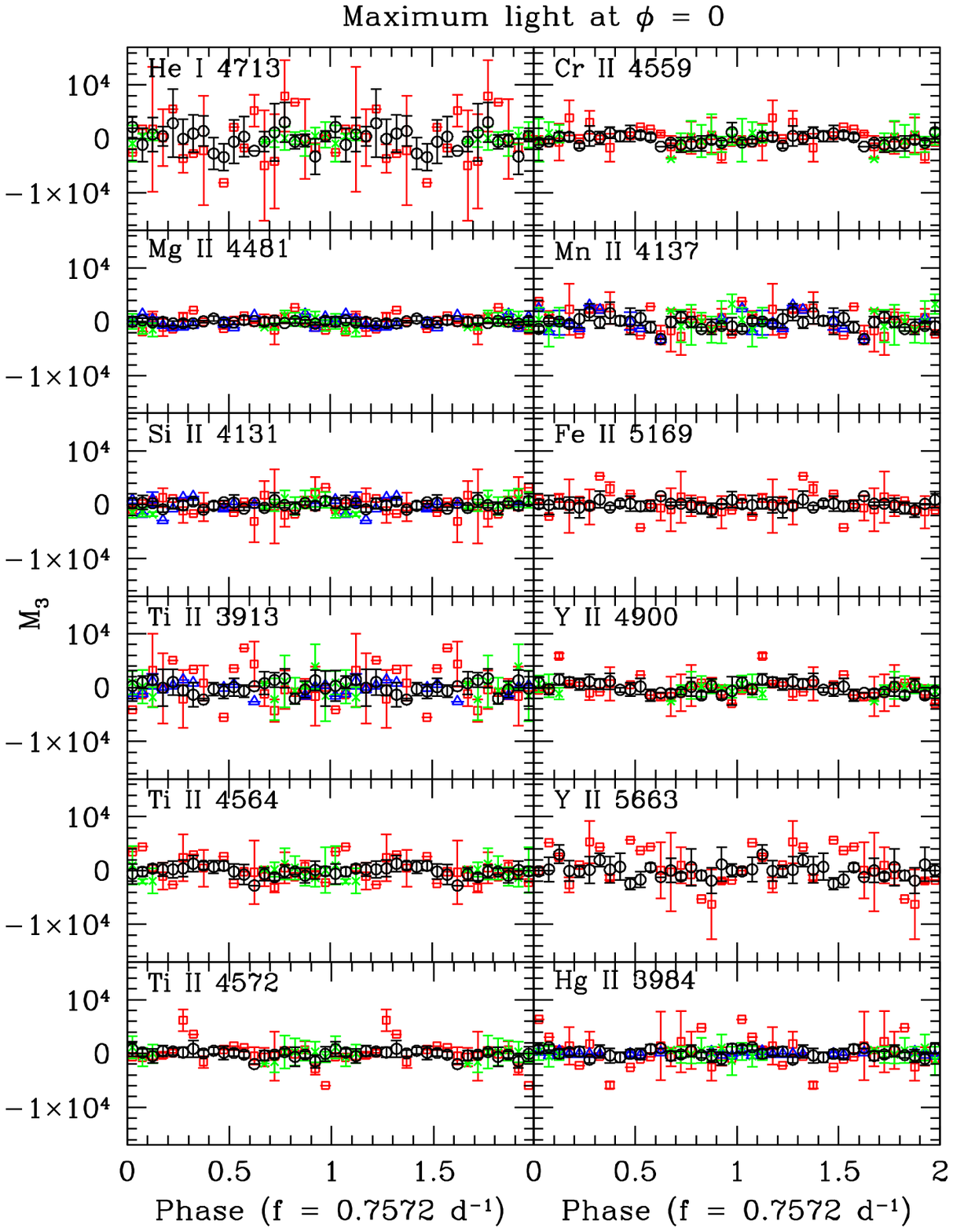}
\end{minipage}
\caption{Variations of the zeroth ({\it top left panels}), first ({\it top right panels}), second ({\it bottom left panels}), and third ({\it bottom right panels}) line-profile moments as a function of phase ({\it circles}: HARPS, {\it squares}: HERMES, {\it triangles}: UVES, and {\it crosses}: SARG data). For the zeroth and second moments, the data shown are normalised to the mean value for a given dataset. For the first and third moments, these are the deviations with respect to the mean value for a given dataset. The data have been grouped into 0.05-phase bins. The uncertainties are the 1-$\sigma$ dispersion of the measurements within a given phase bin. Data points without error bars are based on a single measurement. Phase zero is fixed at maximum light according to the \c \ light curve.} 
\label{fig_moment}
\end{figure*}

Figure \ref{fig_moment} shows the variations in the four moments folded with $f_1$ for all lines. We adopted a zero phase at maximum light (based on the \c \ light curve to occur at HJD 2\,455\,819.514). The phase-locked nature of the variations is much clearer for the HARPS data that are of high quality and that have the best temporal sampling (and also precise radial velocities for what concerns the first moment). The patterns are not as well-defined for the other datasets. For this reason, we have grouped all the data into intervals of 0.05 in phase. The existence of a periodicity is supported by the fact that lines of the same element (\ion{Ti}{ii} $\lambda$4564, 4572 and \ion{Y}{ii} $\lambda$4900, 5663) vary in concert, albeit with more or less clarity depending on the line strength, the possible presence of unrecognised blends, and the robustness of the continuum normalisation. The line-to-line differences in the amplitudes (see, e.g., M$_0$ for \ion{Y}{ii} $\lambda$4900, 5663) have a physical origin and arise, for instance, from a different sensitivity to temperature changes. The large scatter of the \ion{He}{i} $\lambda$4713 measurements may be attributed to the weakness of this line. Variations with a semi amplitude of typically $\sim$5\% in the EWs and 1 \kms \ in the line centroid are observed for the variable features. 

As a second step, we used FAMIAS to perform for the HARPS data a sinusoid fit of the time series of each moment through a least-square minimisation using the expression:
\begin{equation}
F(t) = F_0 + A\sin \left[2\pi \, (f_1 t + \phi)\right],
\label{equation_fit}
\end{equation}
 
\hspace*{-0.6cm} where $F_0$ is the zero point, $A$ is the amplitude, and $\phi$ the phase. The frequency, $f_1$, was held fixed at 0.7572 d$^{-1}$, while these three parameters were adjusted using the Levenberg-Marquardt algorithm. 

\begin{figure}[t]
\centering
\includegraphics[width=9.0cm]{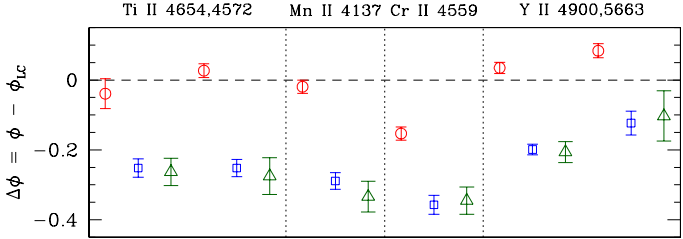}
\caption{Difference between the phase of the fitting functions for the line moments and for the light curve (circles: M$_0$; squares: M$_1$; and triangles: M$_3$).}
\label{fig_phase_shifts}
\end{figure}

For a set of lines for which a periodic signal was clearly detected in their moments, Table \ref{tab_phase_shifts} shows the difference between the phase found and what was derived for the light curve, $\phi_{\rm LC}$, using exactly the same fitting procedure. These results are illustrated in Fig.~\ref{fig_phase_shifts}. As can be seen, M$_1$ and M$_3$ are clearly correlated. This indicates that the centroid variations are primarily due to changes in the line shape and not to a general shift of the line profile as would arise from binary motion. On the other hand, there is a clear phase shift of $\Delta \phi$ $\sim$ 0.25 compared to M$_0$. This is best seen for \ion{Y}{ii} $\lambda$4900 (Fig.~\ref{fig_moment}): while maximum EW corresponds to maximum light, the maxima of the line displacement in velocity occur at $\phi$ $\sim$ 0.25 and 0.75. The increase in the line strength close to maximum brightness is observed for all lines.\footnote{The spectra are normalised to the continuum. Line profiles in spectra taken close to maximum light are therefore expected to be diluted because of the increased continuum level and therefore to appear weaker. However, the variations are expected to be very small based on the amplitude of the \c \ light curve (line strength lower by $\sim$0.3\%). More importantly, the increase in the line strength close to maximum light is contrary to what is expected, and it demonstrates that it arises from an enhanced line strength and is not an artefact of the varying continuum level.} However, noteworthy are the small phase shifts that appear to be present between different lines (compare in particular the behaviour of the Cr and Y lines in Fig.~\ref{fig_phase_shifts}). We return to this point in the following.

\begin{table}[t]
\caption{Difference between the phase of the fitting functions for the line moments and for the light curve.}
\centering
\label{tab_phase_shifts}
\begin{tabular}{l|ccc}
\hline\hline
     & \multicolumn{3}{c}{$\Delta \phi$ = $\phi$ -- $\phi_{\rm LC}$} \\
Line & M$_0$ & M$_1$ & M$_3$\\
\hline
\ion{Ti}{ii} $\lambda$4564 & --0.039\p0.043 & --0.252\p0.026 & --0.263\p0.039 \\
\ion{Ti}{ii} $\lambda$4572 &  +0.027\p0.020 & --0.252\p0.025 & --0.275\p0.053 \\
\ion{Mn}{ii} $\lambda$4137 & --0.019\p0.019 & --0.289\p0.024 & --0.334\p0.044 \\
\ion{Cr}{ii} $\lambda$4559 & --0.153\p0.019 & --0.357\p0.027 & --0.345\p0.039 \\
\ion{Y}{ii} $\lambda$4900  &  +0.035\p0.016 & --0.199\p0.015 & --0.206\p0.030 \\
\ion{Y}{ii} $\lambda$5663  &  +0.084\p0.020 & --0.123\p0.034 & --0.103\p0.072 \\
\hline
\end{tabular}
\end{table}

\begin{figure*}[t]
\begin{minipage}[t]{0.19\textwidth}
\includegraphics[angle=90,width=4\textwidth]{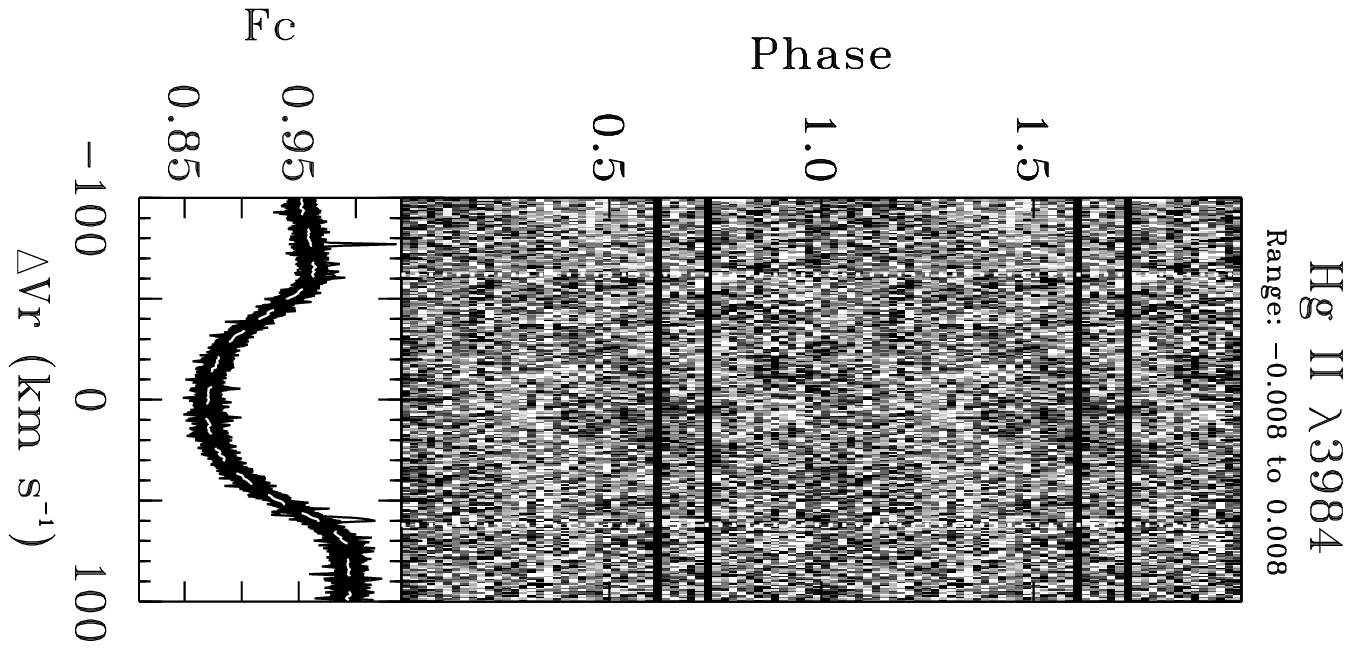}
\end{minipage}
\begin{minipage}[t]{0.19\textwidth}
\includegraphics[angle=90,width=4\textwidth]{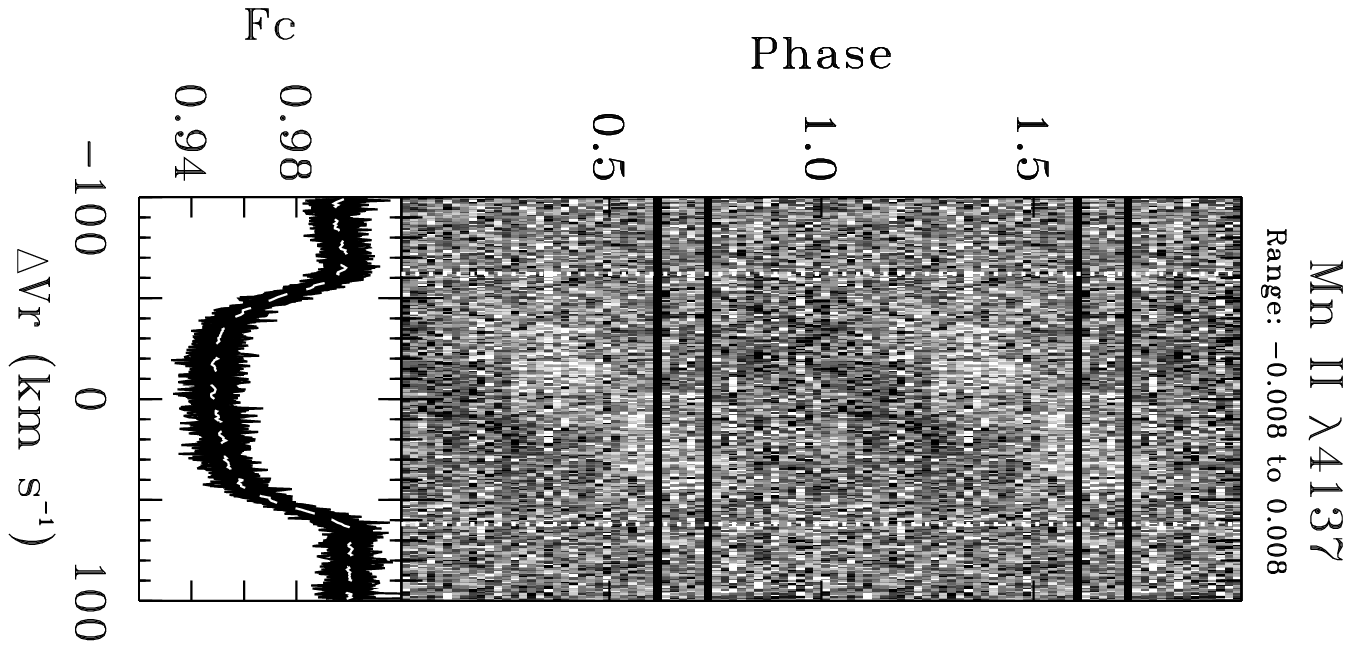}
\end{minipage}
\begin{minipage}[t]{0.19\textwidth}
\includegraphics[angle=90,width=4\textwidth]{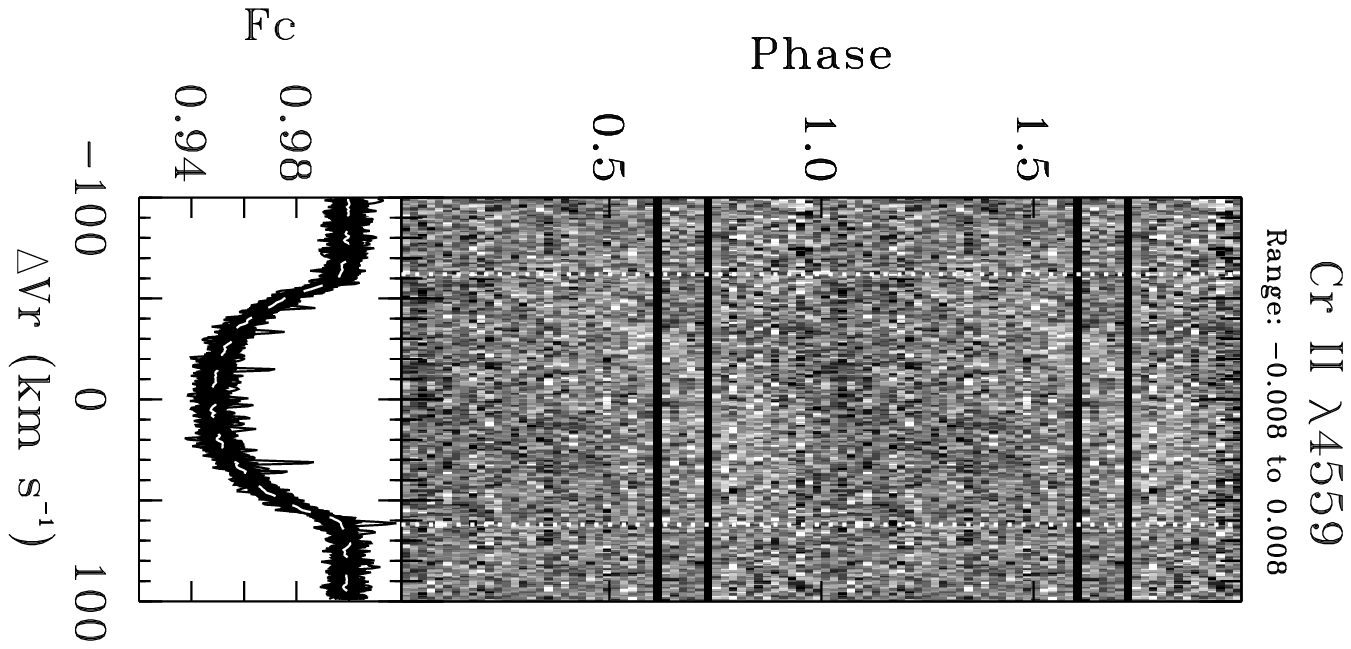}
\end{minipage}
\begin{minipage}[t]{0.19\textwidth}
\includegraphics[angle=90,width=4\textwidth]{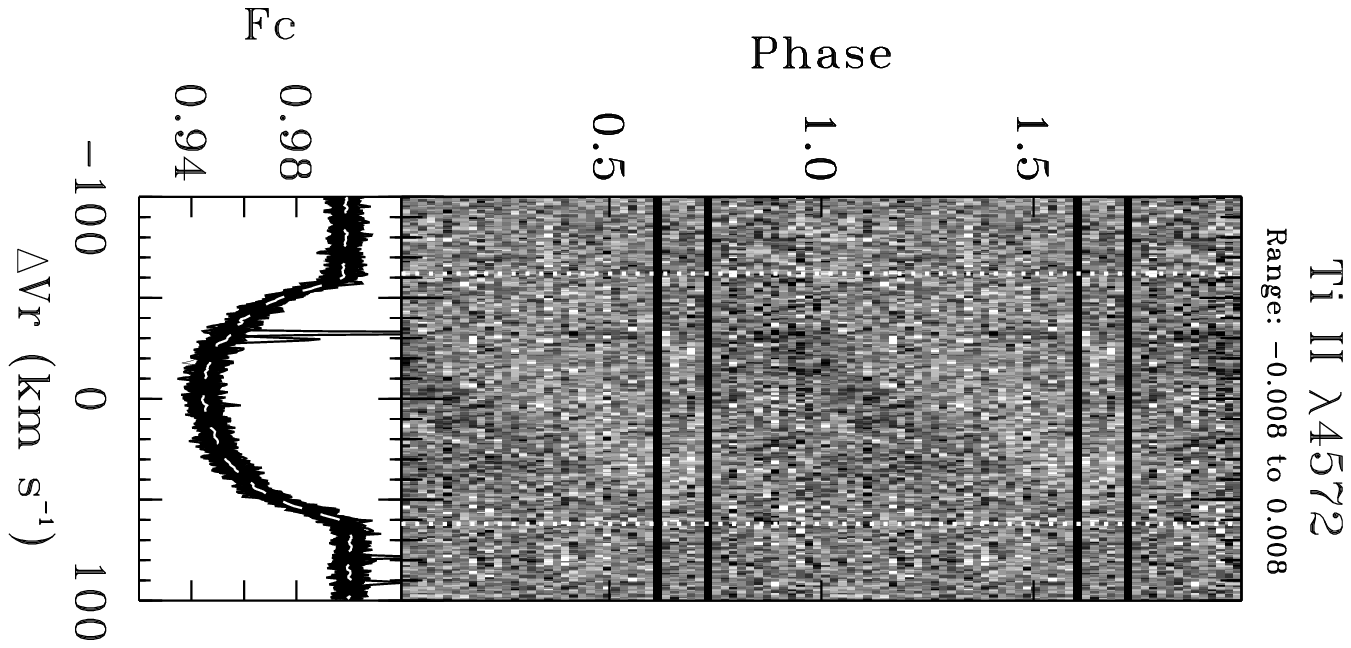}
\end{minipage}
\begin{minipage}[t]{0.19\textwidth}
\includegraphics[angle=90,width=4\textwidth]{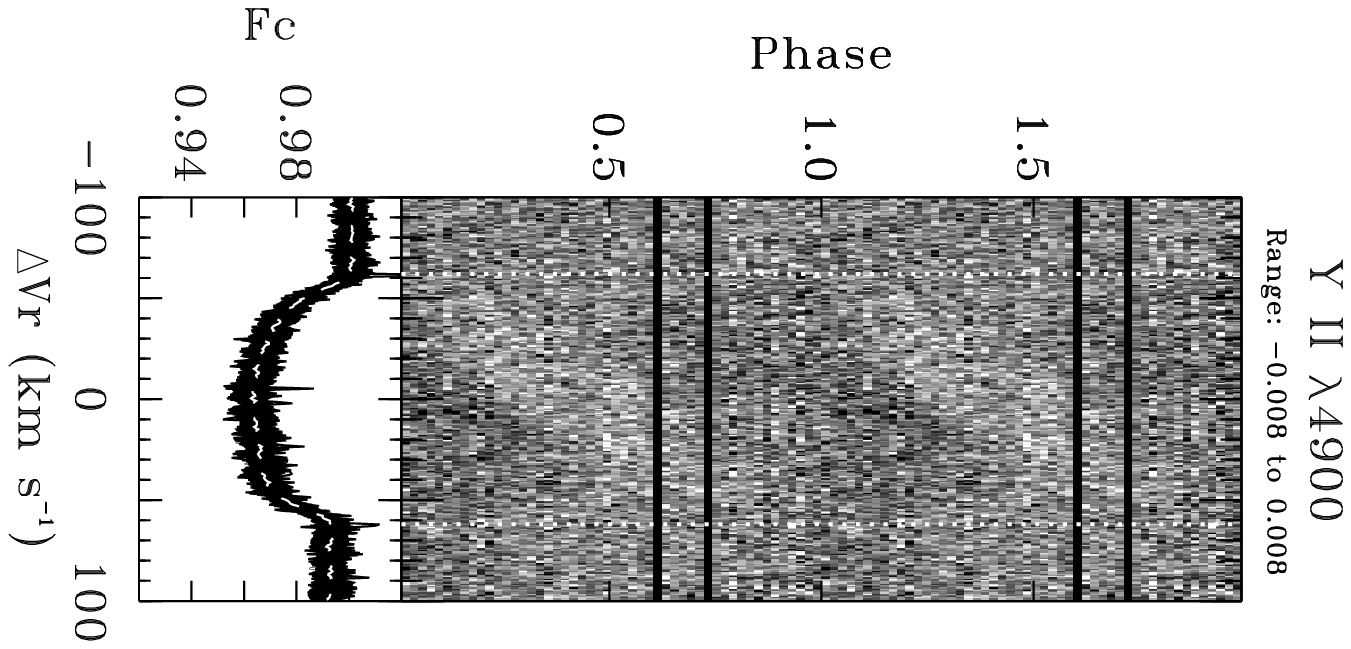}
\end{minipage}
\vskip -1.8cm
\caption{Grey-scale plots of the time series of the residuals for a few selected spectral lines (\ion{Hg}{ii} $\lambda$3984, \ion{Mn}{ii} $\lambda$4137, \ion{Cr}{ii} $\lambda$4559, \ion{Ti}{ii} $\lambda$4572, and \ion{Y}{ii} $\lambda$4900) in the case of the HARPS data. The leftmost panel illustrates the distinct behaviour of \ion{Hg}{ii} $\lambda$3984. (Fig.~\ref{fig_least_squares_pixel1D} shows that the variations are not significant.) These residuals (the mean profile subtracted from the individual profiles) were binned to a 0.02 phase interval. A deficit of absorption appears brighter in these plots. The two vertical dotted lines show the velocities corresponding to \p\vsini \ (determined in Sect.~\ref{sect_abundance_analysis}). The lower panel presents the superposition of the individual profiles (mean profile overplotted as a dashed line). The spectra are displayed in the stellar rest frame. The cosmic-ray events falling in the line profiles have been removed during the period search (Sect.~\ref{sect_lpv}).} 
\label{fig_greyscales}
\end{figure*}

As illustrated in Fig.~\ref{fig_fourier_Y_II_4900}, the average across the line profile of all the Fourier spectra corresponding to the pixel-to-pixel variations shows a significant signal at $f_1$ for some spectral features. Figure \ref{fig_greyscales} shows for the HARPS data grey-scale plots of the variations displayed by lines pertaining to different elements as a function of phase. The variations are very subtle and are almost buried in the noise. However, one can discern an absorption bump (or, alternatively, a dip) moving redwards across the line profiles. In contrast to lines of strongly enhanced elements (Ti, Cr, Mn, and Y), there is no evidence of any variability above the noise level for species with near-solar abundances (He, Si, Mg, and Fe). The peculiar case of Hg is discussed below.

\begin{figure}[h]
\centering
\includegraphics[width=6.75cm]{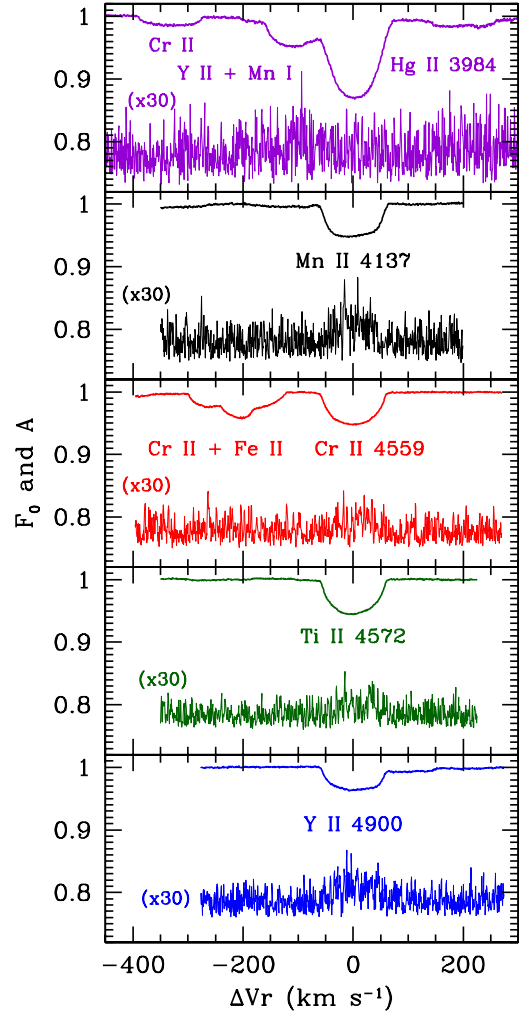}
\caption{Results of the fitting of the variations affecting a given pixel for a region encompassing the lines shown in Fig.~\ref{fig_greyscales}. The zero point and the amplitude of the fitting function are shown in each panel. The amplitude is arbitrarily shifted along the ordinate axis and multiplied by 30 for visibility purposes. To allow for a more robust normalisation of \ion{Hg}{ii} $\lambda$3984, the red wing of H$\epsilon$ was used to define pseudo-continuum regions.}
\label{fig_least_squares_pixel1D}
\end{figure}

\begin{figure}[t]
\centering
\includegraphics[width=7.5cm]{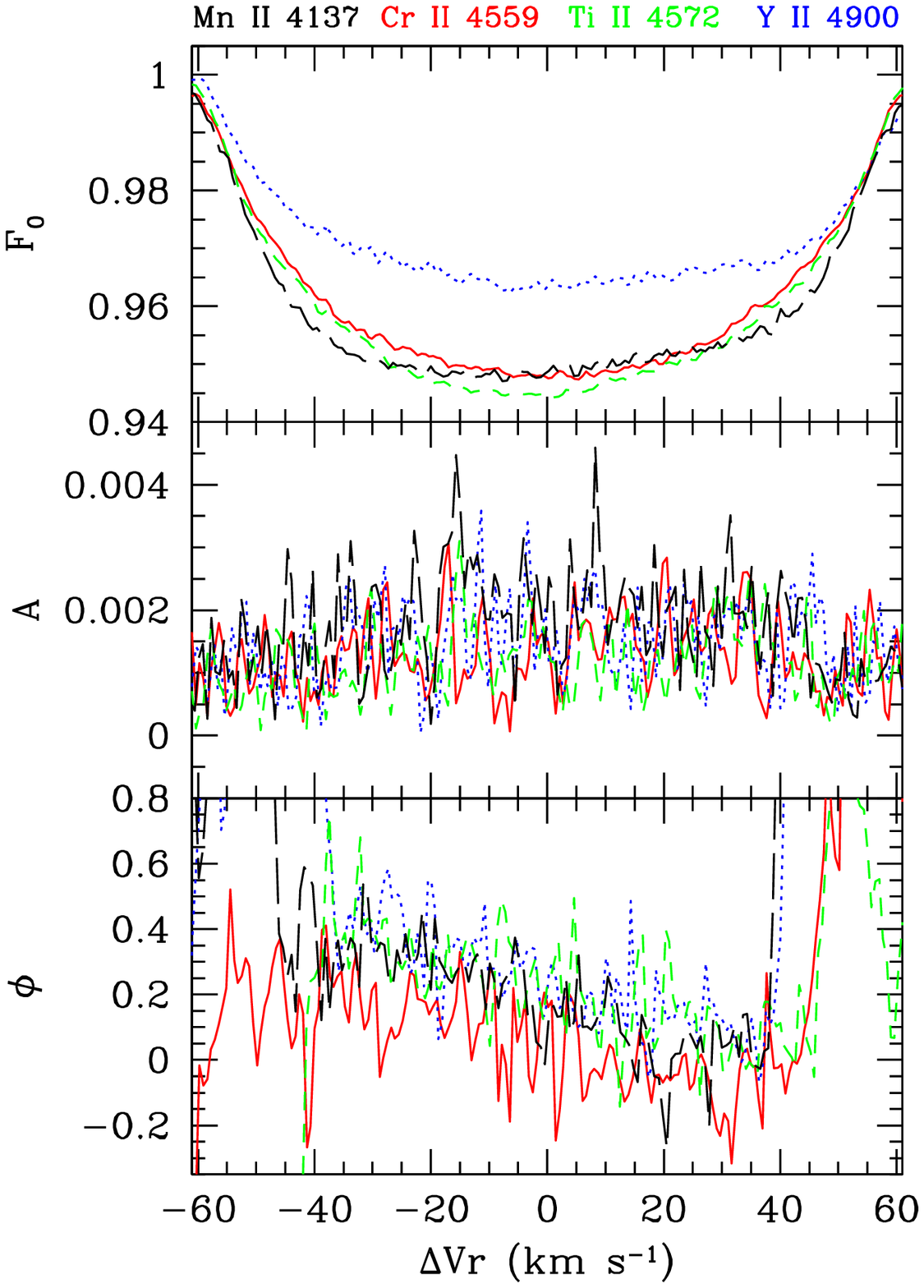}
\caption{Results of the fitting of the variations affecting a given pixel across \ion{Mn}{ii} $\lambda$4137 ({\it long-dashed line}), \ion{Cr}{ii} $\lambda$4559 ({\it solid line}), \ion{Ti}{ii} $\lambda$4572 ({\it short-dashed line}), and \ion{Y}{ii} $\lambda$4900 ({\it dotted line}). From top to bottom: zero point, amplitude, and phase (in units of 2$\pi$).}
\label{fig_phase_shift_pixel1D}
\end{figure}

Similar to what has been done for the moments of the line profiles, we have used Equ.~\ref{equation_fit} to fit the variations affecting a given pixel across the line profiles. For the lines shown in Fig.~\ref{fig_greyscales}, a significant variability compared to neighbouring continuum regions is observed (Fig.~\ref{fig_least_squares_pixel1D}). The notable exception is Hg, which is strongly overabundant, yet no significant variability is detected for the line investigated (\ion{Hg}{ii} $\lambda$3984). As already mentioned for the line moments (Fig.~\ref{fig_phase_shifts}) and also hinted at by a close visual inspection of Fig.~\ref{fig_greyscales}, this analysis reveals a significant phase shift amounting to $\Delta \phi \sim$ 0.15 in the pattern of variability of \ion{Cr}{ii} $\lambda$4559 compared to that of the other species investigated (see bottom panel of Fig.~\ref{fig_phase_shift_pixel1D}). The variations were only confidently detected for velocities (in the stellar rest frame) ranging from about --45 to +45 \kms. As a result, the data shown in Fig.~\ref{fig_phase_shift_pixel1D} for the phase are not reliable outside this range because the fit did not converge.

\begin{figure}[t]
\centering
\includegraphics[width=9.0cm]{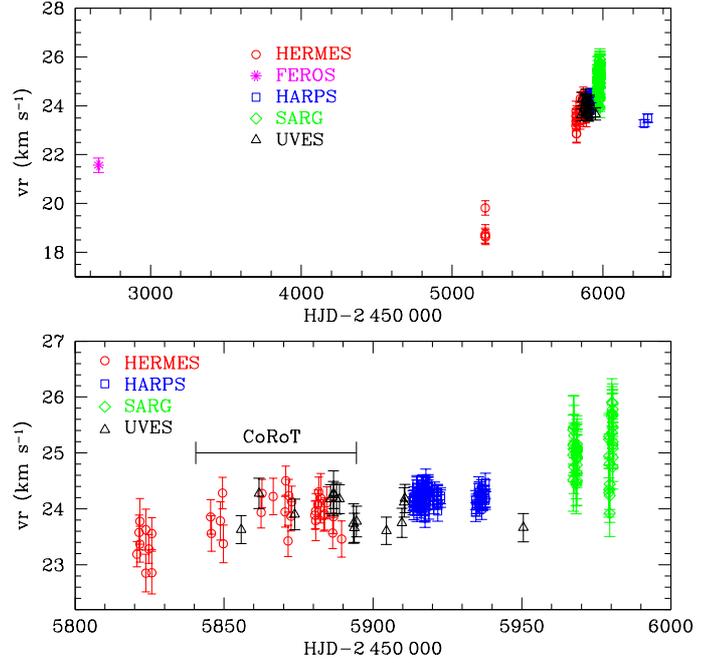}
\caption{Variations in the centroid of \ion{Mg}{ii} $\lambda$4481 \AA \ as a function of the observation date (this line is not covered by the Tautenburg spectra). The upper panel shows the whole dataset, while the lower panel shows the observations obtained close to the \c \ observations.}
\label{fig_radial_velocities}
\end{figure}

Figure \ref{fig_radial_velocities} shows the radial velocities calculated as the centroid (i.e., first moment) of the strong \ion{Mg}{ii} $\lambda$4481 line. This line offers the double advantage of being the strongest metallic feature and the least affected by short-term velocity variations (see Fig.~\ref{fig_moment}). There is clear evidence of long-term (monthly) changes demonstrating that HD 45975 is part of a binary system, as found for many HgMn stars \citep[e.g.,][]{catanzaro04,scholler10,scholler12}. Significant differences are found for HERMES and HARPS spectra secured some months to years apart. This rules out an explanation in terms of instrument-to-instrument zero-point offsets (the fibre entrance aperture projected on the sky is 2.5\arcsec for HERMES and 1.4\arcsec for HARPS. The secondary is thus not the contaminating source discussed in Sect.~\ref{sect_imagettes}). There is also evidence of a slight, gradual increase in the radial velocities by about 1 \kms \ during around the five months during which the star was intensively monitored. We attempted to estimate the parameters of the binary system, but more data are needed to confidently derive the orbital elements. There is no trace of the secondary in the integrated spectrum. The light ratio between the two components is high, so treating HD 45975 as a single star is legitimate when deriving the parameters and abundances (Sect.~\ref{sect_abundance_analysis}).

\section{Discussion}\label{sect_discussion}
Two points are of utmost importance when trying to identify the origin of the photometric and spectroscopic variations observed. First, the level of variability dramatically varies from one species to another. The elements whose lines display periodic variations in their moments or shape are those (with the notable exception of Hg discussed below) that are strongly overabundant with respect to solar (see Fig.~\ref{fig_abundances}): Ti, Cr, Mn, and Y. In contrast, the species with roughly solar abundances show no evidence of any such changes (He, Mg, Si, and Fe). At least for the strong \ion{Mg}{ii} $\lambda$4481 line, these variations are not masked by noise, and tight constraints can be placed on the level of any changes present. 

In the case of classical B-type pulsators, the line-profile variations are essentially due to the surface velocity field that varies with time \cite[e.g.,][]{de_ridder02}. Similar variations affecting lines of all species, whether they are enhanced in the photosphere or not, would therefore be expected. However, the situation may be different in case of a strongly chemically stratified atmosphere. For the roAp stars, for instance, the short wavelength of the pulsation wave in the vertical direction, coupled with the line-formation regions being located at very different heights in the atmosphere, leads to orders of magnitude difference between the amplitudes of the different ions. The level of variability generally increases towards the upper atmospheric layers as a result of the decreasing density \citep[e.g.,][]{kochukhov2008,khomenko09}. One may expect a different response of the line profiles in the case of the SPB-like pulsations that could be present in HgMn stars (much longer pulsation periods, hence wavelength). 

However, the lack of variability observed in HD 45975 for mercury (Figs.~\ref{fig_moment} and \ref{fig_least_squares_pixel1D}) is difficult to reconcile with this interpretation, since it would require that the \ion{Hg}{ii} line-formation region is very different from that of the other ions displaying an overabundance. This is at odds with the predictions of non-magnetic diffusion models, which indicate that Ti, Cr, Mn, and Hg should all accumulate\footnote{Unpublished models by G. Alecian (private communication) also show the same behaviour for Y.} in the upper atmosphere layers for a HgMn star with a \teff \ similar to that of HD 45975 (\citealt{alecian10}; see also \citealt{leblanc09}). We notice, however, that the run of the abundances as a function of depth in the photosphere differs from one element to another, since the buildup of the stratifications, which is non-linear and time-dependent, depends on the atomic properties of each element \citep[as discussed by][]{alecian11}. In the context of abundance spots, the lack of variability for Hg in HD 45975 might be attributed to geometrical effects (implying that the spot morphology differs from that of the other species) and/or a distribution of mercury at the surface that is more uniform than for Ti, Cr, Mn, and Y. The \ion{Hg}{ii} $\lambda$3984 line is variable in most HgMn stars, but there are exceptions \citep[\object{66 Eri A};][]{makaganiuk11b}. Conversely, there are cases where the variations are dramatic for the Hg lines, but are (if any) below the detection limit for lines of other elements \citep[\object{$\alpha$ And};][]{wade06}. 

Second, a phase shift amounting to $\Delta \phi \sim$ 0.15 (or about 5 hours) is observed between the patterns of variability of Cr and that of Ti, Mn, and Y (Figs.~\ref{fig_phase_shifts}, \ref{fig_greyscales}, and \ref{fig_phase_shift_pixel1D}). Such a phenomenon (known as the ``Van Hoof'' effect) is observed in classical pulsators with a high-amplitude dominant mode and is interpreted as a running shock wave propagating through the atmosphere and reaching the line-formation regions of the various ions at different times. This leads to time lags of minutes for stars without very extended atmospheres, such as $\delta$ Scuti \citep[e.g.,][]{mathias97} or $\beta$ Cephei \citep[e.g.,][]{mathias93} stars. 

Perhaps more relevant to the present case is to discuss the observations of the roAp stars where the pulsation wave propagates outwards through a strongly chemically stratified medium (keeping in mind that these stars possess very strong magnetic fields unlike the HgMn stars). The running (magneto)acoustic wave is found to travel at a velocity comparable to the sound speed in the outer atmosphere, and the time lags are once again much shorter than observed here \citep[a few minutes, e.g.,][]{ryabchikova07}. Based on these timescale considerations, it is therefore very unlikely that the time lags observed in HD 45975 can be attributed to a running pulsation wave. Furthermore, an additional requirement that is hard to justify on theoretical grounds would be that the \ion{Cr}{ii} line-formation region is spatially distinct from that of \ion{Ti}{ii}, \ion{Mn}{ii}, and \ion{Y}{ii}. One may speculate that the phase lag exhibited by \ion{Cr}{ii} $\lambda$4559 is instead due to chromium spots that are separated in longitude by about 60 degrees compared to those of titanium, manganese, and yttrium. 

The spectral pattern of variability consists in an excess absorption component moving from blue to red across the line profile. In addition, the EW and centroid variations are shifted in phase by (about) one quarter of a cycle, and maximum EW occurs close to (or at) maximum light. This is also more naturally explained by patches at the stellar surface producing excess light when they pass through the central meridian and creating a component with excess absorption travelling across the line profile as they are carried by stellar rotation. The period found (1.321 d) is compatible with the rotational one according to our \vsini \ and radius estimates. Specifically, assuming \vsini \ = 61\p2 \kms \ and $R$ = 1.8\p0.5 R$_{\odot}$ (Sect.~\ref{sect_abundance_analysis}) restricts the inclination angle with respect to the line of sight, $i$, to be higher than about 45$\degr$. A star seen near equator-on is consistent with its \vsini \ being rather high for a HgMn star \citep{smith96}.

There is mounting evidence that spatially-extended abundance spots can exist in the photospheres of HgMn stars. A patchy surface distribution of some chemical species has first been firmly established from the detection of large, daily changes in the strength of \ion{Hg}{ii} $\lambda$3984 in the primary component of \object{$\alpha$ And} \citep{adelman02}, but has also been subsequently found in other HgMn stars \citep[e.g.,][]{kochukhov05,briquet10,hubrig11}. To date, Doppler imaging techniques have been used to infer the distribution (and evolution) of chemical spots in four B stars with HgMn peculiarity: \object{AR Aur A} \citep{hubrig06,savanov09,hubrig10,hubrig11,hubrig12}, \object{66 Eri A} \citep{makaganiuk11b}, \object{$\alpha$ And A} \citep{adelman02,kochukhovetal07}, and \object{$\phi$ Phe} \citep{makaganiuk12,korhonen13}. All the stars discussed here are slow or moderately fast rotators with equatorial velocities below 55 \kms \ and rotation periods in the range 2--10 days. The diversity of behaviours observed and the very limited number of objects investigated in detail obviously do not allow us to anticipate what spot morphology might be expected for HD 45975. Furthermore, it should be kept in mind that the spot distribution may change on a monthly/yearly timescale \citep{kochukhovetal07,hubrig10,hubrig12,korhonen13}, and that the presence of a close companion might have a profound effect on the spot morphology \citep[e.g.,][]{hubrig06,makaganiuk11b}. However, the sinusoidal character of the photometric and spectroscopic changes suggest a rather simple spot configuration at the time of the observations. That the lines of Ti, Mn, and Y vary in phase indicates that the abundance spots of these species lie at roughly the same stellar longitude, as found for various elements in other HgMn stars \citep{makaganiuk11b,makaganiuk12,korhonen13}.

In stars with strong, organised magnetic fields and with diffusion processes at work (e.g., the classical Ap/Bp stars), the transport of the ionised species along the field lines is amplified and leads to a large-scale inhomogeneous distribution of some elements at the surface. The formation of abundance spots in HgMn stars might be related to the existence of weak magnetic fields \citep[e.g.,][]{alecian12}, but such fields still need to be detected. Spectropolarimetric observations have not been attempted so far for HD 45975, and it is therefore unclear how the spots are created in the first place and then sustained.

\section{Conclusions}\label{sect_conclusions}
Although our data are too sparse to constrain the binary orbit, our measurements show that HD 45975 is part of a single-lined binary system with a period of months to years. We have presented evidence of monoperiodic variations (with a period of 1.321 d) in the photometric and spectroscopic datasets. The amplitude of the variations observed in the \c \ light curve is comparable to what has previously been reported in other HgMn stars from space observations \citep{alecian09,balona11}. On the other hand, the level of line-profile variability is lower than observed for most HgMn stars \citep[see, however, the case of \object{$\mu$ Lep};][]{kochukhov11}. 

These changes are more compatible with the rotational modulation of abundance spots on the surface. Their existence has already been firmly established in other HgMn stars. As discussed above, a spotted photosphere can be more easily reconciled with: (i) the level of spectral variability which varies widely from one element to the next and is below our detection limit for mercury and (ii) the phase lag between the patterns of variability presented by different ions. These arguments rest to a large extent on our current theoretical understanding of the photospheres of these stars, so more realistic calculations are needed to put our conclusions on a firmer footing. Developing physical models of a spotted HgMn star to reproduce the line-profile and photometric changes observed is also required to address the validity of our hypothesis. If we accept this interpretation, our observations place tight upper limits on the amplitude of any variations arising from pulsational instabilities in HD 45975 and suggest that they are unlikely to produce photometric variations above the $\sim$50 ppm level (see Fig.~\ref{fig_fourier_corot}). Such stringent observational constraints may lead to better understanding of the excitation and damping of pulsation modes in HgMn stars \citep{turcotte_richard03,alecian09}.

\begin{acknowledgements}
TM acknowledges financial support from Belspo for contract PRODEX GAIA-DPAC. EN acknowledges support from the NCN grant 2011/01/B/ST9/05448. Calculations were carried out in Wroc\l aw Centre for Networking and Supercomputing ({\tt http://www.wcss.wroc.pl}), Grant No. 214. MR acknowledges financial support from the FP7 project {\it SPACEINN: Exploitation of Space Data for Innovative Helio- and Asteroseismology}. EP, FB, and MS acknowledge financial support from the PRIN-INAF 2010 ({\it Asteroseismology: looking inside the stars with space- and ground-based observations}). CU sincerely thanks the South African National Research Foundation (NRF) for awarding innovation post-doctoral fellowship, Grant No. 73446, and NRF MULTI-WAVELENGTH ASTRONOMY RESEARCH PROGRAMME (MWGR), Grant No: 86563, Reference: MWA1203150687. PIP acknowledges funding from the European Research Council under the European Community's Seventh Seventh Framework Programme (FP7/2007-2013)/ERC grant agreement n$^\circ$227224 (PROSPERITY) and from the Belgian federal science policy office BELSPO (C90309: CoRoT Data Exploitation). This paper uses some of the observations made at the South African Astronomical Observatory (SAAO). We wish to thank ESO for the allocation of Director's Discretionary Time. The observations with TNG/SARG have been funded by the Optical Infrared Coordination network (OPTICON), a major international collaboration supported by the Research Infrastructures Programme of the European Commissions Sixth Framework Programme. We are grateful to Wojciech Borczyk, Tomasz Kowalczyk, Adrian Kruszewski, Krystian Kurzawa, and Anna Przybyszewska for their assistance with the observations. We also wish to thank Marc Ollivier (IAS) for his help with the imagettes. This work made use of GAUDI, the data archive and access system of the ground-based asteroseismology programme of the \c \ mission. The GAUDI system is maintained at LAEFF. LAEFF is part of the Space Science Division of INTA. We would like to thank the anonymous referee for a comprehensive report and valuable comments. TM wishes to thank the colleagues from the ASTA team of the department of Astrophysics, Geophysics and Oceanography of the University of Li\`ege for enlightening discussions. The spectroscopic data were analysed with the software package FAMIAS developed in the framework of the FP6 European Coordination Action HELAS. This research made use of NASA's Astrophysics Data System Bibliographic Services and the SIMBAD database operated at the CDS, Strasbourg (France).
\end{acknowledgements}

\end{document}